\documentclass[lettersize,journal]{IEEEtran}
\usepackage{cite,citesort}
\usepackage{graphicx,graphics}
\usepackage{amssymb,amsfonts,array,mdwmath,mdwtab}
\usepackage[cmex10]{amsmath}
\usepackage{enumerate}
\usepackage{multirow}
\usepackage{color,comment}
\usepackage{textcomp}
\usepackage{url}

\usepackage{algorithmic}
\usepackage{algorithm}

\makeatletter
\let\MYcaption\@makecaption
\makeatother
\usepackage[font=footnotesize]{subcaption}
\makeatletter
\let\@makecaption\MYcaption
\makeatother

\begin{document}

\title{
High-Capacity and Low-PAPR BICM-OFDM Systems
Using Non-Equiprobable and Non-Uniform 
Constellation Shaping
With Clipping and Filtering
}

\author{Eito Kurihara,~\IEEEmembership{Student Member,~IEEE},
and Hideki Ochiai,~\IEEEmembership{Fellow,~IEEE}
\thanks{%
The authors are with the 
Graduate School of Engineering, The University of Osaka, Osaka 565-0871, 
Japan.
(email: kurihara-eito@ieee.org, 
ochiai@comm.eng.osaka-u.ac.jp)}
}

\maketitle

\begin{abstract}
We address a design of high-capacity and low-peak-to-average power ratio~(PAPR) 
orthogonal frequency-division multiplexing~(OFDM) systems based on
bit-interleaved coded modulation (BICM) utilizing 
non-equiprobable and non-uniform~(NENU) 
constellations as well as clipping and filtering~(CAF).
The proposed constellations are generated using a truncated Gaussian distribution and the merging of constellation points, where the former creates a non-uniform constellation~(NUC), 
and the latter 
adjusts
the number of signal points 
for further improving the 
total
bit-wise mutual information (BMI).
Unlike other 
exhaustive search-based approaches, 
the proposed constellations are uniquely determined by only two parameters associated with NUC and cardinality. Due to this property of limited degrees of freedom,
the complexity required for the numerical optimization process can be significantly low.
We focus on the constellation design based on one dimension,
i.e., pulse amplitude modulation~(PAM), which facilitates the reduction
of demapping complexity for the BICM receiver.
The use of CAF at the transmitter can efficiently reduce the PAPR of OFDM signals; however, it introduces clipping noise that may degrade error rate performance, making the application of clipping noise cancellation~(CNC) at the receiver essential.
Therefore, we optimize the NENU constellations in the presence of CAF and CNC.
Simulation results demonstrate that the combination of constellation shaping with CAF and CNC enables BICM-OFDM systems to simultaneously achieve low PAPR and high spectral efficiency
over additive white Gaussian noise (AWGN) as well as frequency-selective 
fading channels.
Furthermore, comparative studies confirm that the proposed system 
significantly outperforms the single-carrier counterpart (i.e., DFT-precoded BICM-OFDM) 
in terms of PAPR and bit error rate~(BER) performance over fading channels.
\end{abstract}

\section{Introduction}
\label{sec:Introduction}

To achieve significantly high spectral efficiency under limited spectral resources,
the use of quadrature amplitude modulation (QAM) with very large constellation sizes,
together with capacity-approaching channel codes, is essential.
A common practical approach is to employ bit-interleaved coded modulation (BICM),
which separates modulation and coding by decomposing the transmitted constellation
into parallel binary-input channels, to which conventional binary channel codes can
readily be applied.
When combined with standard QAM, 
the constellation 
can be decomposed into
two 
Gray-labeled and equidistant (i.e., uniform) pulse amplitude modulation (PAM) constellations,
each corresponding to the in-phase (I) and quadrature (Q) components.
This structure significantly simplifies the complexity of demapping and decoding
processes at the receiver~\cite{szczecinski2015bit}.

A key challenge in
BICM, especially with  
standard 
PAM,
is its achievable information rates that
are inferior to the Shannon limit. 
The standard 
PAM constellation, with each constellation point spaced equally from its neighbors, differs significantly from Gaussian signals that achieve the Shannon limit. This discrepancy leads to a noticeable gap, especially when aiming for a 
high target spectral efficiency.
To reduce this gap, constellation shaping, which essentially tailors the distribution of PAM constellations
to approximate a Gaussian distribution, plays an important role.

Constellation shaping is categorized as geometric and probabilistic shaping.
Probabilistic shaping adjusts the probability distribution of the uniform PAM, 
whereas geometric shaping modifies the placement of constellation points~\cite{steiner2017comparison,qu2019probabilistic}.
Probabilistic shaping is typically implemented by introducing a distribution matcher, which entails high computational and storage complexity at the transmitter~\cite{schulte2015constant}.
In contrast, 
geometric shaping is simply achieved by altering the locations of constellation points, making it compatible with current systems~\cite{fay2016overview,singya2021survey}.
Accordingly, 
the design of a geometrically shaped constellation, 
often referred to as a non-uniform constellation~(NUC), 
represents a significant area of ongoing research, particularly in conjunction with
BICM~\cite{sillekens2022high,gumucs2023simplified,chen2022geometrically}.

In view of practical applications to systems with high spectral efficiency,
we focus on the 
design 
of NUCs based on 
PAM (i.e., one-dimensional NUCs) 
with large constellation sizes.
It is well known that two-dimensional
constellations can achieve 
better
performance in terms of the gap from the Shannon limit.
However, 
their implementation becomes increasingly challenging as the target information rate increases
due to the prohibitively high demapping complexity 
associated with large constellation sizes. 
In fact,
existing wireless communication systems targeting very high spectral efficiency often
employ modulation schemes based on one-dimensional constellations.
For example, 
the Advanced Television Systems Committee's third-generation standard~(ATSC~3.0) 
adopts only one-dimensional NUCs for
higher modulation orders such as $1024$ and $4096$~\cite{atsc30}.

A major drawback of 
constellation shaping is
the increase in peak-to-average power ratio (PAPR), as it aims to reduce the average power for a given maximum amplitude.
This leads to
a significant reduction of power efficiency at the transmitter,
as linear amplification of high PAPR signals is challenging without sacrificing 
power amplifier efficiency~\cite{ochiai2013analysis}.
Nevertheless, 
the majority of modern wireless communication systems adopt 
orthogonal frequency-division multiplexing (OFDM) 
because of its advantages, such as high spectral efficiency and robustness against 
fading channels.
Since OFDM transmits multiple signals simultaneously, its PAPR 
becomes
high 
irrespective 
of the constellations used for each subcarrier, with or without shaping~\cite{ochiai_TCOM_2001}. 
In such systems, it is necessary to apply PAPR reduction techniques originally developed for OFDM 
while ensuring high compatibility with constellation shaping.

\subsection{Our Contributions}

In this work, our target is to implement 
a BICM-OFDM system that achieves both high spectral efficiency and low PAPR
without imposing high complexity at the transmitter side.

In our shaping approach, 
a Gray-labeled non-uniform PAM constellation, 
whose cardinality is a power of two, 
is first systematically constructed based on a truncated Gaussian distribution characterized 
by a single parameter. The number of constellation points is then sequentially reduced by merging the closest points.
This merging process improves the achievable information rate at a given signal-to-noise ratio (SNR),
while reducing the demapping complexity.
As a result, some constellation points are selected with higher probability than the others, leading to a non-equiprobable constellation. We refer to the resulting constellations as
{\em non-equiprobable and non-uniform}~(NENU) constellations.

For achieving low PAPR, 
we adopt clipping and filtering~(CAF)~\cite{ochiai2002performance}
among various PAPR reduction techniques proposed for OFDM.
While many PAPR reduction schemes impose demanding complexity at the transmitter side,
CAF can be implemented in a straightforward manner.
Furthermore, it is directly applicable to BICM-OFDM systems with any constellation shaping.
Therefore, combining constellation shaping with CAF is a natural consequence for BICM-OFDM.
However, CAF inevitably causes performance degradation due to the in-band nonlinear distortion called clipping noise. Therefore, we also consider the application of clipping noise cancellation~(CNC)~\cite{chen2003iterative,sun2021,wachowiak2023clipping} 
to mitigate this impairment at the receiver.
To tailor the constellation design to OFDM transmission
under CAF and CNC, the two shaping parameters (i.e., the 
truncation level associated with the initial Gaussian distribution
and the number of distinct signal points) 
are jointly optimized 
by taking into account the effects of both
additive white Gaussian noise (AWGN) and residual clipping noise. 
Overall, the proposed system can simultaneously achieve low PAPR and high spectral efficiency, even within a practical framework of OFDM over frequency-selective fading channels.

We summarize the main contributions of this work as follows:
\begin{itemize}
    \item We propose a novel design algorithm for 
NENU PAM constellations, which combines a geometric design based on a truncated Gaussian distribution with a rate-adaptive constellation merging scheme\footnote{%
In~\cite{kurihara2025systematic}, we first presented the concept of merging adjacent signal points for constellations designed according to truncated Gaussian distributions. 
The algorithm presented in this work is a refined version of~\cite{kurihara2025systematic}
in that it can enhance the BMI through the merging process.
}. 
Numerical evaluations demonstrate that the proposed scheme achieves performance comparable to the numerically optimized NUC adopted in the ATSC standard.

    \item We design BICM-OFDM systems with low PAPR and high spectral efficiency, where low PAPR is achieved by CAF at the transmitter and high spectral efficiency is supported by CNC at the receiver that is compatible with our 
    proposed shaping scheme.
    Furthermore, the NENU constellations are optimized in the presence of CAF and CNC.

    \item Through computer simulations, 
we compare 
the proposed BICM-OFDM systems with 
their single-carrier counterparts~\cite{myung2006single}
based on discrete Fourier transform~(DFT) precoding
that employs the same constellation shaping in terms of
the complementary cumulative distribution function~(CCDF) of PAPR,
demonstrating that the proposed system with the proper clipping ratio 
can achieve lower PAPR than the single-carrier systems.

\item Comparison of the resulting bit error rate~(BER) performance over 
AWGN
and frequency-selective fading channels shows
that the proposed BICM-OFDM system provides noticeable gains in both channels.
This is due to the CNC that successfully compensates for the BER degradation caused by CAF 
with no additional 
complexity
associated with 
the use of
constellation shaping.
It outperforms the single-carrier system with the same constellation over fading channels even with lower PAPR. 
\end{itemize}

\subsection{Related Work}

Several systematic geometric shaping approaches based on Gaussian-like 
PAM
constellations have been proposed in the literature~\cite{koike2017bit,wiens2020constellation,boutros2018geometric} 
to enhance the 
achievable information rate 
while keeping design and implementation complexity modest. 
Despite their respective design variations, these methods share the common design principle of partitioning a Gaussian distribution into equiprobable 
subregions and assigning a representative constellation point to each subregion, 
as
originally suggested in~\cite{sun1993approaching}.
In our previous work~\cite{kurihara2024design}, we proposed a systematic approach to designing non-uniform 
PAM
constellations using a truncated Gaussian distribution adjusted by a single parameter. 
This method generates the PAM constellation, whose distribution can be adjusted
ranging from Gaussian-like to uniform,
and we identify the parameter that maximizes the achievable information rate  
under specified channel SNRs.
Nevertheless, the designed constellations fail to achieve shaping gains 
as high as those of 
ATSC~3.0~\cite{atsc30}, which rely on the optimization through an 
exhaustive numerical search.
Moreover, the underlying design principles of such optimized NUCs remain largely unexplained, limiting theoretical understanding and systematic reproducibility.

As a low-complexity implementation of probabilistic shaping,
a many-to-one mapping approach that does not rely on a distribution matcher
can be employed~\cite{yankov2014rate,chang2013symbol,zhou2020enhancement}. 
This approach introduces redundant bit labels, which are assigned to a smaller number of constellation points such that the resulting probability mass function (PMF) approximates a Maxwell–Boltzmann (i.e., quantized Gaussian) distribution.
Extensions to relatively high-order QAM constellations have recently been investigated in~\cite{yankov2016constellation,jia2023many,cao2020probabilistic}.
Since these approaches only require modifications to the bit-to-symbol mapping table, their implementation cost is significantly lower than those based on a distribution matcher.
However, they tend to result in lower shaping gains due to their coarse approximation of the target distribution unless a considerably large number of shaping bits are used to adjust the PMF.
Recent works~\cite{li2024model,soleimanzade2023hybrid,stark2019joint,maneekut2020hybrid} have also explored the integration of geometric and probabilistic constellation shaping. Many of these approaches design the constellation geometry and symbol probabilities separately through numerical optimization.

In contrast, our approach offers a unified framework that jointly determines both the positions of constellation points and their probabilities based on a systematic design principle.
We also address the optimization of constellations in the presence of CAF and CNC, which involve complex nonlinear signal processing operations at both the transmitter and receiver.
Most existing constellation shaping approaches through numerical optimization 
deal with 
analytically tractable channels such as AWGN. On the other hand,
since the proposed approach relies on a predefined family of constellations 
characterized by 
limited degrees of freedom,
it facilitates efficient optimization even 
for such channels that cannot be described by explicit mathematical models.
Lastly, the application of 
geometric
shaping to OFDM systems in conjunction with PAPR reduction techniques has not been well
investigated in the literature.\footnote{%
Trellis shaping~\cite{forney92} can be applied to PAPR and average power reduction of BICM-OFDM systems~\cite{ochiai_comm04,yoshizawa2015}.
However, the transmitter complexity required for effective PAPR reduction is 
considerably higher than CAF, especially when OFDM has a large number of subcarriers.}

\subsection{Organization}

The rest of this paper is organized as follows:
Section~\ref{sec:Shaping} introduces our NENU constellation construction algorithm, followed by the performance analysis with parameter optimization in Section~\ref{sec:performance}.
Section~\ref{sec:System} describes the proposed BICM-OFDM system model employing CAF and CNC, 
where the
achievable information rate
in this system model is analyzed.
Simulation results are presented in Section~\ref{sec:Result}, focusing on the BER performance 
over AWGN and practical fading channels as well as the CCDF of the resulting PAPR.
Finally, Section~\ref{sec:Conclusion} concludes this work.

\subsection{Notation}
For a given scalar $A_k$, the corresponding bold font $\mathbf{A}$ denotes its row-vector form.
Sets of real positive numbers, complex numbers, and natural numbers are given by ${\mathbb R}_+$, ${\mathbb C}$, and ${\mathbb N}$, respectively,
whereas ${\mathbb F}_2$ represents the binary field.
Calligraphic fonts such as ${\mathcal A}$ denote sets, with their cardinality given by $|{\mathcal A}|$, whereas
${\cal N}\left(\mu,\sigma^2\right)$
and
${\cal CN}\left(\mu,\sigma^2\right)$
represent the real-valued Gaussian distribution
and the circularly symmetric complex Gaussian distribution, respectively,
both having mean $\mu$ and variance $\sigma^2$.
The expectation operator is represented by $\mathbb{E}\left[\cdot \right]$.

\section{A New Constellation Shaping Algorithm}
\label{sec:Shaping}

The proposed design algorithm for NENU constellations consists of two primary steps.
First, an initial Gray-labeled PAM constellation is derived by partitioning a truncated Gaussian distribution into equiprobable subregions, where each constellation point
is assigned according to the midpoint 
cumulative distribution function~(CDF) value of the corresponding region.
The resulting PAM distribution is controlled by a single parameter that determines the truncation level of the reference Gaussian distribution.
Next, adjacent subregions are sequentially merged to adjust the number of distinct signal points.
In that case, the reference Gaussian distribution is partitioned such that the 
probability associated with a given subregion is proportional to the 
number of assigned binary labels.
This algorithm facilitates a flexible constellation design in terms of both the number of constellation points and their distribution, making it suitable for rate-adaptive BICM systems.
We demonstrate that when the two design parameters (i.e., the number of distinct signal points and the Gaussian truncation level) are jointly optimized, the resulting constellations closely match the optimized NUCs adopted in the ATSC standard. 
Unlike the latter, which requires a comprehensive table to specify the constellation points, 
only a pair of parameters uniquely determines our constellation.
As described in Section~\ref{sec:System}, this ``limited degrees of freedom'' description
turns out to be particularly 
useful when designing constellations over highly complex
channels that may only be modeled by simulation-based approaches.

\subsection{Geometric Design Based on Truncated Gaussian CDF}

We first review an $M$-ary PAM constellation design based 
on the truncated Gaussian distribution~\cite{kurihara2024design},
where each PAM symbol has 
a binary label ${\bf b} \in {\mathbb F}_2^m$ with $m=\log_2 M$.
Due to the symmetry of the constellation, it is sufficient
to consider the constellation points within the positive region.
Let $\mathcal {M}_+
\triangleq \{1, 2, \cdots, M/2\}$ denote a set of constellation indices in this region.
We define the set of corresponding constellation points as
 $\mathcal{A}_{+} \triangleq \{a_k \in {\mathbb R}_+ \,|\, k \in {\mathcal M}_+\}$, 
with the elements ordered in the ascending order, i.e., $0 < a_1 \leq a_2 \leq \cdots \leq a_{M/2}$.
The constellation points are labeled by the binary reflected Gray code (BRGC)
of $m$ bits, where 
the adjacent points differ only by a single bit.
The most significant bit (MSB) is assigned as the sign bit
such that $a_k$ and $-a_k$ share the identical bit labels except for the MSB.
Without loss of generality, we assume that MSB is $0$ for the elements
of $\mathcal{A}_{+}$. The resulting constellation, including the negative region, is
given by $\mathcal{A} = \mathcal{A}_{+} \cup \mathcal{A}_{-}$, where
$\mathcal{A}_{-} \triangleq \{-a_k  \,|\, k \in {\mathcal M}_+\}$.

Since each PAM constellation point is selected with an equal probability,
for a given target probability distribution,
their locations are determined such that 
their CDF
values are evenly distributed.
Specifically, since the initial target distribution is Gaussian,
we consider a random variable $X \sim {\cal N}\left(0,\frac{1}{2}\right)$.
The CDF of $X$ is given by
\begin{equation}
F_X(x)=\frac{1+\operatorname{erf}(x)}{2},
\label{eq:Fx0}
\end{equation}
where $\operatorname{erf}(x) = \frac{2}{\sqrt{\pi}}\int_{0}^{x}e^{-t^2}dt$ is the error function.
Since the range of $X$ is unbounded, we truncate its range within $[-b, b]$ 
for a given parameter $b>0$.
Accordingly, the range of the CDF in the positive amplitude region
should be modified as 
$\left[\frac{1}{2}, F_X(b)\right]$. 
Then, 
the tentative value $\hat{a}_k$,
which corresponds to the point $a_k$ before energy normalization,
is determined 
by the following equality:
\begin{equation}
F_X(\hat{a}_k) = \frac{1}{2} +l_k \frac{F_X(b) - \frac{1}{2}}{M} ,
\label{eq:Fx}
\end{equation}
where
\begin{equation}
l_k = 2k - 1, \quad \,\text{ for }\, k  \in \mathcal {M}_+
\label{eq:index}
\end{equation}
is the integer value unique to the $k$th constellation point.
Let us define $\rho$ as
\begin{equation}
\rho \triangleq \operatorname{erf}(b), \quad \rho\in(0,1]
\label{eq:parameter}
\end{equation}
such that the CDF 
in the positive range of amplitude
reduces to
$\left[\frac{1}{2}, \frac{1 + \rho}{2}\right]$ according to \eqref{eq:Fx0}. 
We may then express $\hat{a}_k$ in~\eqref{eq:Fx} using~\eqref{eq:Fx0} 
as
\begin{align}
\hat{a}_k= \phi_M(l_k; \rho) \triangleq
\operatorname{erf}^{-1}\left(
\rho
 \frac{l_k}{M}
\right),
\label{eq:mapper}
\end{align}
where $\phi_M: \mathbb{N} \to \mathbb{R}_+$ is a constellation mapper 
for a given parameter $\rho$.
Finally, scaling the magnitude of $\hat{a}_k$ in accordance with
the energy constraint 
leads to the desired constellation point $a_k$:
\begin{equation}
a_k = \sqrt{\frac{E_s}{2}}
\frac{
\hat{a}_k}
{\sqrt{\frac{2}{M}
\sum_{\ell =1}^{M/2} \hat{a}^2_\ell}
},
\label{eq:norm}
\end{equation}
where $E_s$ denotes the average transmit energy per complex-valued symbol.
In our subsequent numerical calculation of the constellation points, 
we set $E_s = 1$ for simplicity.

The above design algorithm uniquely determines the non-uniform PAM
constellation using a single parameter 
$\rho \in (0,1]$ for a given modulation order $M$.
When $\rho = 1$ (i.e., without truncation), the resulting constellation coincides with the Gaussian-like PAM proposed in~\cite{boutros2018geometric}.
In contrast, 
we may express the Maclaurin series of $F_X(x)$ up to the first order as
\begin{align}
F_X(x) \approx \frac{1}{2} + \frac{1}{\sqrt{\pi}} x ,
\nonumber
\end{align}
which indicates that $F_X(x)$ linearly increases (i.e.,
$X$ becomes a uniform distribution)
for $X \in [-b, b]$ in the limit of  $b \to 0$ ($\rho \to 0$).
For this reason, we denote $\rho=0$ as the standard uniform PAM in what follows.

\begin{figure}[tbp]
\begin{center}

  \begin{minipage}{\linewidth}
    \centering
    \includegraphics[width=\hsize]{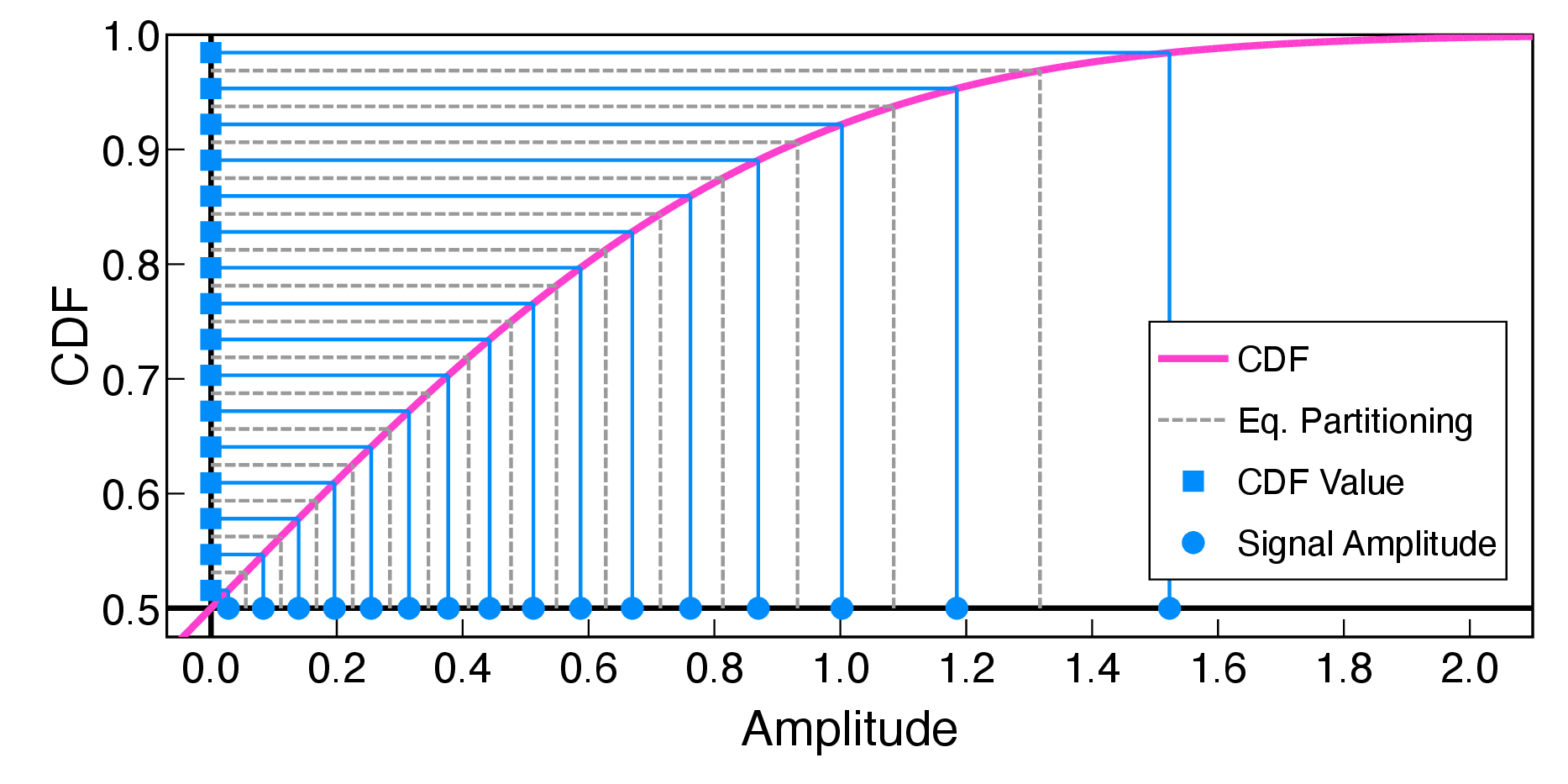} 
    \footnotesize (a) $\rho=1.0$
  \end{minipage}

  \vspace{0.6em}

  \begin{minipage}{\linewidth}
    \centering
    \includegraphics[width=\hsize]{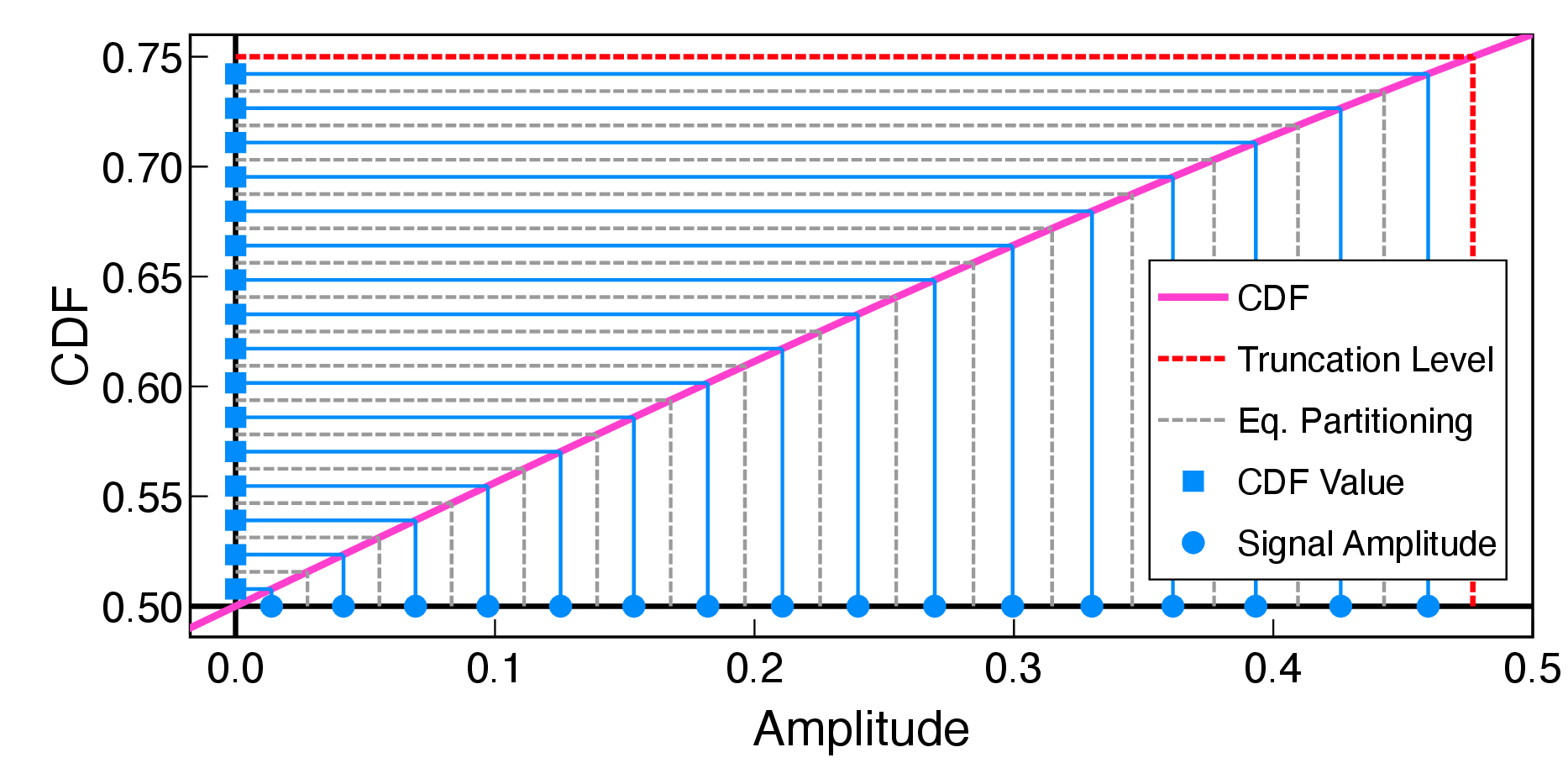}
    \footnotesize (b) $\rho=0.5$
  \end{minipage}

  \caption{Constellation point selection based on equiprobable partitioning of the Gaussian CDF with $M=32$. Due to the symmetry, only the positive region is shown. (a) is the case with $\rho=1.0$ (i.e., without truncation), while (b) corresponds to the case with $\rho=0.5$}.
  \label{fig:Method1}
  
\vspace{-0.5em}
\end{center}
\end{figure}

The relationship between the CDF values and the selected constellation points $\{\hat{a}_k\}$
for the case of $M=32$ and $\rho = 1$ (i.e., without truncation) 
is illustrated in Fig.~\ref{fig:Method1}(a).
The values $\phi_M(2(k-1);\rho)$ and $\phi_M(2k;\rho)$ 
serve as the boundaries of the $k$th equiprobable subregion, 
which are depicted as the corresponding gray dashed lines. 
Each subregion is assigned a single binary label ${\bf b} \in \mathbb{F}_2^m$,
and the representative constellation point 
$\hat{a}_k$ is then selected 
to match the midpoint CDF value (i.e.,  $\phi_M(l_k;\rho)$) of the corresponding region.
On the other hand, the case of $\rho=0.5$ 
is shown in 
Fig.~\ref{fig:Method1}(b), where the effective range of the original Gaussian CDF is now 
truncated to 
$\left[\frac{1}{2}, \frac{1 + \rho}{2}\right]= [0.5, 0.75]$,
as indicated by the red dashed line. 
Within this truncated range, we observe that the CDF increases almost linearly,
and thus the resulting constellation points become more uniformly spaced, approaching those of the conventional uniform PAM constellation.

\subsection{Constellation Point Merging Process}
\label{sec:proposedShaping}

In coded modulation systems that do not utilize bit-wise interleaving, the achievable information rate, referred to as the constellation-constrained capacity, consistently increases with higher-order modulations.
In BICM systems, the sum of the bit-wise mutual information (BMI) serves as the achievable information rate, and
the constellation size that maximizes the total BMI depends on the channel SNR, indicating that
high-order constellations may become suboptimal in low-SNR regimes~\cite{caire1998bit,zollner2013optimization}.
Therefore, we next consider dynamically adjusting the number of constellation points 
by incorporating a strategy of merging constellation points
into the truncated Gaussian-based geometrical design described above.

We define $N$ as the number of distinct points in an $M$-ary PAM symbol with $N \leq M$,
which indicates that a PAM constellation with $M$ distinct labels
has only $N$ distinct constellation points.
We denote this constellation as $(M, N)$-PAM in this work. 
The construction of $(M, N)$-PAM is given by the following recursive principle: 
adjacent point pairs 
in $\hat{\mathcal{A}}_+ \triangleq \{\hat{a}_k\,|\, k \in {\mathcal{M}_+}\}$ 
with the minimum Euclidean distance are merged sequentially until the number of distinct points reaches the specified size $N$.
The merging of the two symmetrical pairs in the positive and negative regions occurs simultaneously, leading to a reduction of two constellation points at each merging step. 
Therefore, for an initial modulation order of $M$, the procedure can be executed a maximum of 
$M/2-1$ times in principle, ultimately leading to the standard BPSK constellation (when $N=2$). 

Each merging operation combines the corresponding subregions into a new, unified subregion, and the representative constellation point is placed 
to match
the midpoint CDF value of this modified subregion.
Now, suppose that for a given parameter $\rho$, the constellation points $\hat{a}_i = \phi_M(l_i;\rho)$ and $\hat{a}_j = \phi_M(l_j;\rho)$ (with $i\neq j$) have the minimum Euclidean distance 
at a certain merging step.
Then, these two points are merged into a single point by updating their indices and 
amplitude values as\footnote{
Since the initial indices $\{l_k\}$ are set to be positive odd numbers
according to \eqref{eq:index}, the first merging step
always applies to the two positive odd numbers. As a result, 
the merged index $l^{\prime}$ is guaranteed
to be a natural number.
}
\begin{equation}
\hat{a}_i = \hat{a}_j := \phi_M\left(
l^{\prime};\rho \right), \quad l^{\prime} := \frac{l_i + l_j}{2}.
\label{eq:mapper2}
\end{equation}
After this process, the multiple binary labels originally associated with $\hat{a}_i$ and $\hat{a}_j$ are mapped to the identical signal amplitude.
The proposed shaping algorithm, which is summarized 
in {\bf Algorithm~\ref{algorithm}},
applies to a specific pair of initial modulation order $M$ and the number of 
distinct points $N$ after the merging process, with 
$\mathcal{L}$ 
representing the set of the constellation indices in the positive 
amplitude region that have unique constellation point locations. The algorithm thus 
terminates when $|\mathcal{L}| = N/2$.

In what follows, 
a set of the resulting 
distinct
PAM constellation points
located in both positive and negative 
regions, 
after normalization through~\eqref{eq:norm},
will be denoted by ${\cal A}\left(M,N,\rho\right)$,
which is uniquely determined
by the two parameters, $N$ and $\rho$,
for a given modulation order $M$.
Furthermore, since each binary label is converted to a corresponding PAM symbol,
we define the symbol mapping function as
\begin{align}
\varphi:  {\mathbb F}_2^m \to {\cal A}\left(M,N,\rho\right)
\label{eq:mapping}
\end{align}
such that $\varphi(\bf b)$ represents the PAM symbol that has the binary label 
of length $m$ specified by ${\bf b}$. We note that if $N < M$
the mapping function $\varphi$ will be characterized by many-to-one mapping.

\begin{figure}[tbp]
    \begin{center}
    \includegraphics[width=\hsize]{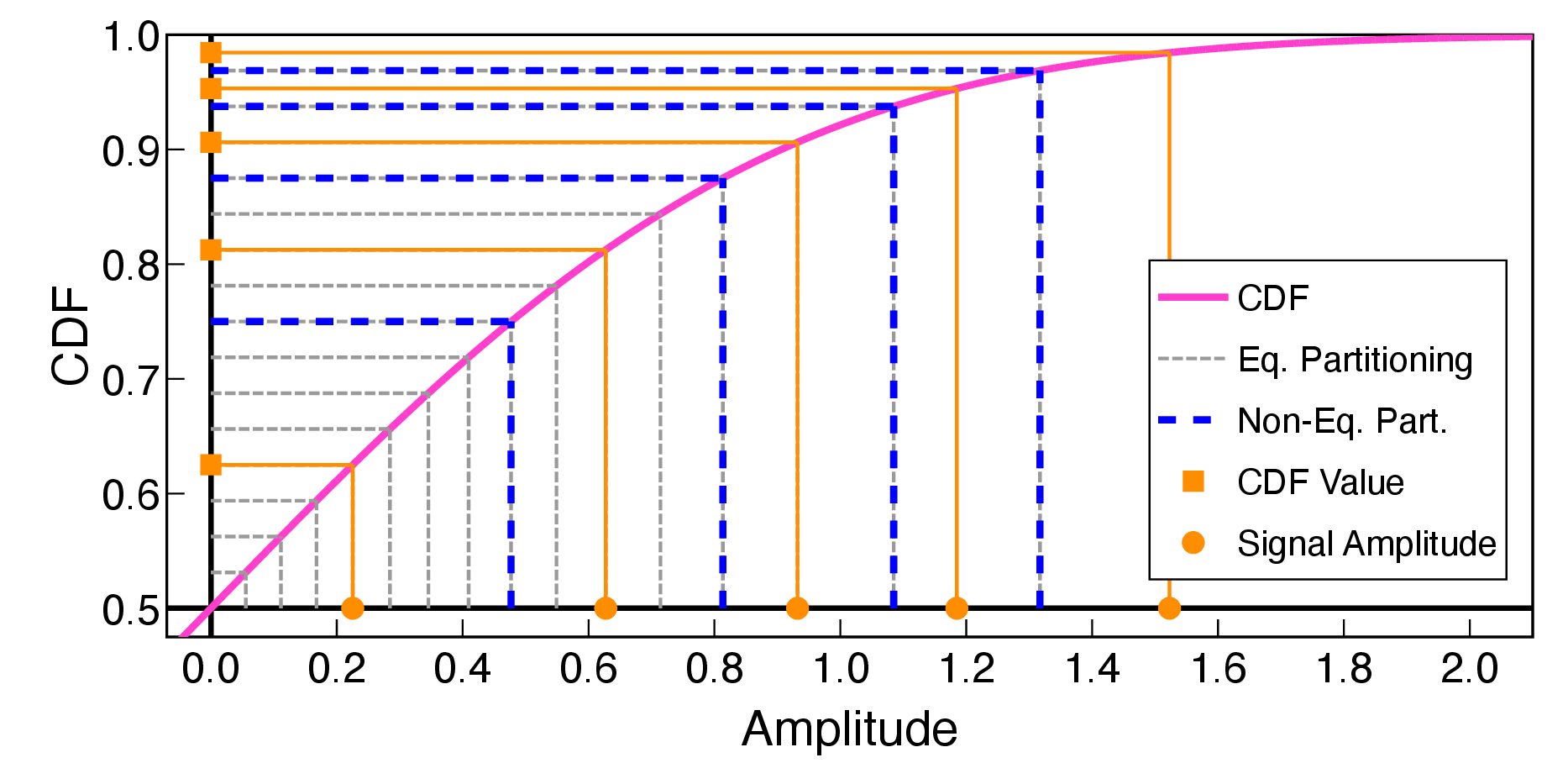}\\
     \caption{Constellation point selection based on non-equiprobable partitioning of the Gaussian CDF ($\rho=1.0$) with $M=32$ and $N=10$.}
\label{fig:Method2}
\end{center}
\vspace{-0.5em}
\end{figure}

\renewcommand{\algorithmicrequire}{\textbf{Input:}}
\renewcommand{\algorithmicensure}{\textbf{Output:}}

\begin{algorithm}[t]
\caption{Construction of $(M,N)$-PAM} 
\label{algorithm}
\begin{algorithmic}[1]

\REQUIRE Modulation order $M$; The number of distinct points $N$; Shaping parameter $\rho \in (0, 1]$
\ENSURE Constellation set $\hat{\mathcal{A}}_+$ 
\vspace{0.5em}

\STATE $\mathcal{L} \gets \{ l_k := 2k-1 \,|\, k = 1, 2, \ldots, M/2 \}$
\STATE $\hat{\mathcal{A}}_+ \gets \{ \hat{a}_k := \phi_M(l_k;\rho) \,|\, k = 1, 2, \ldots, M/2 \}$ \\
\WHILE{$M>N$} 
    \STATE $(i, j) \gets \arg\min_{(i,j)} |\hat{a}_i - \hat{a}_j|$ for  $1 \leq i < j \leq M/2$ \\ 
    \STATE $\mathcal{L} \gets \mathcal{L} \setminus  \{l_i, l_j\}$
    \STATE $\hat{\mathcal{A}}_+ \gets \hat{\mathcal{A}}_+ \setminus  \{\hat{a}_i, \hat{a}_j\}$
    \STATE $l^{\prime} \gets \frac{l_i + l_j}{2}$
    \STATE $\hat{a}_k \gets \phi_M\left(l^{\prime};\rho\right)$
    \STATE $\mathcal{L} \gets \mathcal{L} \cup \{l^{\prime}\}$
    \STATE $\hat{\mathcal{A}}_+ \gets \hat{\mathcal{A}}_+ \cup \{\hat{a}_k\}$
    \STATE $M \gets M - 2$
 \ENDWHILE
\RETURN $\hat{\mathcal{A}}_+$
\end{algorithmic}
\end{algorithm}

\subsection{Example}

\begin{table*}[t]
\begin{center}
\caption{
The $l_k$ indices of constellation points $\hat{\mathcal{A}}_+$ for the $(32, N)$-PAM with $\rho=1$ (no truncation). The second row indicates the binary labels. 
The initial amplitude values are shown in the third row,
while the values after the merging process are shown in brackets.
The number in the leftmost column represents $N$ (distinct points).
}
    \begin{tabular}{|c||c|c|c|c|c|c|c|c|c|c|c|c|c|c|c|c|}
    \hline 
    $k$ & 1 & 2 & 3 & 4 & 5 & 6 & 7 & 8 & 9 & 10 & 11 & 12 & 13 & 14 & 15 & 16\\ \hline
\!\text{label}\!    & \!00000\! & \!00001\! & \!00011\! & \!00010\! & \!00110\! & \!00111\! & 
\!00101\! & \!00100\! & \!01100\! & \!01101\! & \!01111\! & \!01110\! 
& \!01010\! & \!01011\! & \!01001\! & \!01000\! \\ \hline 
    $\hat{a}_k$ & .028 & .083 & .139 & .196 & 
.255 & .315 & .377 & .443 & .512 & .587 & 
.669 & .762 & .870 & 1.003 & 1.185 &  1.523
    \\ \hline
     \hline
     32 & 1 & 3 & 5 & 7 & 9 & 11 & 13 & 15 & 17 & 19 & 21 &23 &25 & 27 & 29 & 31 \\ \hline
     30 & \multicolumn{2}{c|}{2~[.055]} & 5 & 7 & 9 & 11 & 13 & 15 & 17 & 19 & 21 & 23 & 25 & 27 & 29 & 31 \\ \hline
     28 & \multicolumn{2}{c|}{2} &  \multicolumn{2}{c|}{6~[.168]} & 9 & 11 & 13 & 15 & 17 & 19& 21 & 23 & 25 & 27 & 29 & 31 \\ \hline
     26 & \multicolumn{2}{c|}{2} &  \multicolumn{2}{c|}{6} &  \multicolumn{2}{c|}{10~[.284]} & 13 & 15 & 17 & 19 & 21 & 23 & 25 & 27 & 29 & 31 \\ \hline
     24 & \multicolumn{2}{c|}{2} & \multicolumn{2}{c|}{6} & \multicolumn{2}{c|}{10} & \multicolumn{2}{c|}{14~[.410]} & 17& 19 & 21 & 23 & 25 & 27 & 29 & 31 \\ \hline
     22 & \multicolumn{2}{c|}{2} & \multicolumn{2}{c|}{6} & \multicolumn{2}{c|}{10} & \multicolumn{2}{c|}{14} & \multicolumn{2}{c|}{18~[.549]} & 21 & 23 & 25 & 27 & 29 & 31 \\ \hline
    20 & \multicolumn{2}{c|}{2} & \multicolumn{2}{c|}{6} & \multicolumn{2}{c|}{10} & \multicolumn{2}{c|}{14} & \multicolumn{2}{c|}{18} & \multicolumn{2}{c|}{22~[.714]} & 25 & 27 & 29 & 31 \\ \hline
    18 & \multicolumn{4}{c|}{4~[.111]} & \multicolumn{2}{c|}{10} & \multicolumn{2}{c|}{14} & \multicolumn{2}{c|}{18} & \multicolumn{2}{c|}{22} & 25 & 27 & 29 & 31 \\ \hline
    16 & \multicolumn{4}{c|}{4} & \multicolumn{4}{c|}{12~[.346]} & \multicolumn{2}{c|}{18} & \multicolumn{2}{c|}{22} & 25 & 27 & 29 & 31 \\ \hline
    14 & \multicolumn{4}{c|}{4} & \multicolumn{4}{c|}{12} & \multicolumn{2}{c|}{18} & \multicolumn{2}{c|}{22} & \multicolumn{2}{c|}{26~[.932]} & 29 & 31 \\ \hline
    12 & \multicolumn{4}{c|}{4} & \multicolumn{4}{c|}{12} & \multicolumn{4}{c|}{20~[.627]} & \multicolumn{2}{c|}{26} & 29 & 31 \\ \hline
    10 & \multicolumn{8}{c|}{8~[.225]} & \multicolumn{4}{c|}{20} & \multicolumn{2}{c|}{26} & 29 & 31 \\ \hline
    8 & \multicolumn{8}{c|}{8} & \multicolumn{4}{c|}{20} & \multicolumn{2}{c|}{26} & \multicolumn{2}{c|}{30~[1.317]} \\ \hline
    6 & \multicolumn{8}{c|}{8} & \multicolumn{4}{c|}{20} & \multicolumn{4}{c|}{28~[1.085]} \\ \hline
    4 & \multicolumn{8}{c|}{8} & \multicolumn{8}{c|}{24~[.813]} \\ \hline 
    2 & \multicolumn{16}{c|}{16~[.477]} \\ \hline
 
    \end{tabular}  
\label{tab:merger}
\end{center}
\end{table*}

The integer representations $\{l_k\}$ of the constellation points for the positive region $\hat{\mathcal A}_+$ generated by {\bf Algorithm~\ref{algorithm}}
are listed in Table~\ref{tab:merger}
in the case of $(32, N)$-PAM starting from $N=32$ to $2$, where
the initial constellation points are derived from the original Gaussian distribution (i.e., $\rho=1$).
Since
the binary labels are based on BRGC,
the bit labels of the negative region $-a_k$ 
are identical to that of $a_k$ with the MSB replaced by $1$.
Focusing on the first merging step from $N=32$ to $30$, the initial constellation points near the origin (e.g., $\hat{a}_1,\hat{a}_2,\hat{a}_3,\hat{a}_4$) exhibit very small Euclidean distances. 
When the adjacent point pair $\hat{a}_1$ and $\hat{a}_2$ is merged, they are modified into a single constellation point that is assigned two distinct binary labels.
Then, these merged symbols become indistinguishable from each other, while they become more separable from the remaining symbols (e.g., $\hat{a}_3,\hat{a}_4$).
As a result, the constellation behaves similarly to a lower-order constellation with fewer but more reliable symbols.

Fig.~\ref{fig:Method2} illustrates the constellation point selection process 
based on 
the proposed merging scheme
applied to the resulting constellation of Fig.~\ref{fig:Method1}(a), 
with the number of distinct points set to $N = 10$.
As shown in the figure, the original equiprobable subregions (gray dashed lines) are merged into larger subregions, depicted as blue dashed lines.
Each representative constellation point is then selected in the same manner as that in Fig.~\ref{fig:Method1},
but based on the midpoint CDF value of the modified subregion.
This merging process corresponds to a non-equiprobable partitioning of the reference Gaussian distribution, according to the number of binary labels assigned to each constellation point.

Considering the fact that the resulting constellation exhibits a non-uniform 
probability mass function (PMF),
the proposed design can be viewed as a hybrid of geometric and probabilistic shaping.
However, the non-uniformity in symbol probabilities arises from the rearrangement
(i.e., geometrical design) of the Gray-labeled $M$ symbols, where multiple symbols 
are merged to have an identical amplitude. 
Thus, unlike conventional probabilistic shaping schemes that require a distribution matcher, the proposed shaping scheme imposes only the update of
the bit-to-symbol mapping table in a PAM-based BICM system 
with the same modulation order $M$.

\subsection{Comparison With NUC of ATSC~3.0}

\begin{figure}[tbp]
\begin{center}
    \includegraphics[width=\hsize]{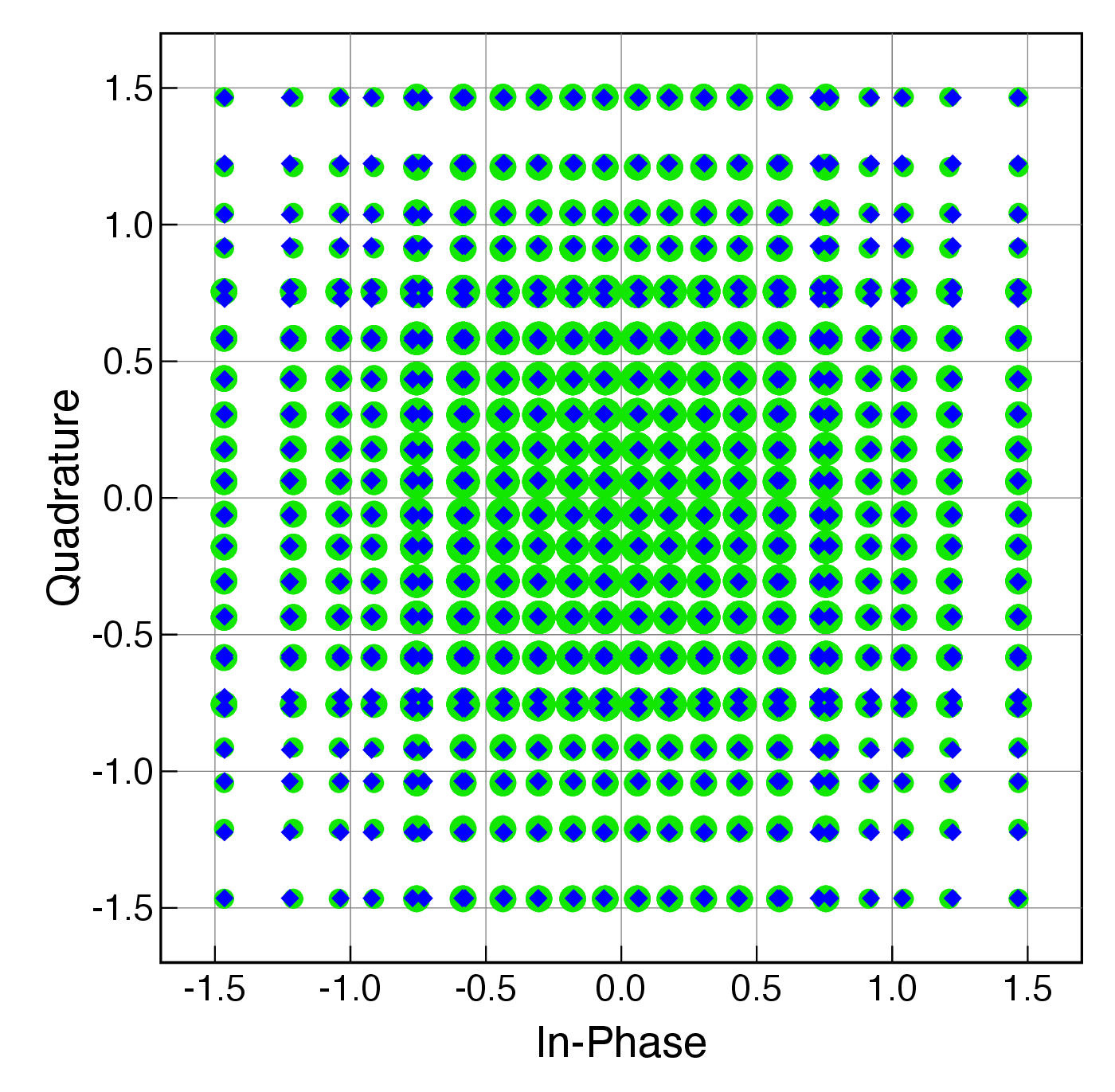}
\end{center}         
    \caption{Comparison between the proposed QAM consisting of two (32,20)-PAM constellations (shown as green circles) and the equivalent one-dimensional NUC adopted in ATSC~3.0 (shown as blue diamonds). Both constellations are optimized for BICM with an information rate of 6.0 bits per symbol (code rate of $3/5$). The size of the circles reflects their probability of occurrence.}
    \label{fig:constellation}
\vspace{-0.5em}
\end{figure}

The proposed 
NENU
constellation consisting of two $(32, 20)$-PAM constellations is depicted using green circles in Fig.~\ref{fig:constellation}, where symbols with higher probabilities are plotted with larger marks. 
The parameter $\rho$ is optimized 
based on
the BMI
for a target spectral efficiency of $6.0$ bits per symbol, corresponding to a code rate of $3/5$.
For comparison, the equivalent $1024$-ary 
one-dimensional NUC adopted in ATSC~3.0 is depicted with blue diamonds. 
While the ATSC NUC individually optimizes all
PAM amplitudes,
many points near the origin are almost overlapping, resembling a smaller number of distinct points.
The proposed constellation, which is systematically derived by {\bf Algorithm~\ref{algorithm}} and
governed by only the two design parameters $N$ and $\rho$, exhibits a remarkably similar configuration to the NUC of ATSC~3.0.
The performance analysis and optimization of these parameters leading to the above result are 
described in the following section.

\section{Performance Analysis}
\label{sec:performance}

This section analyzes the performance that can be achieved by the proposed constellation design through parameter optimization, focusing on achievable gain and demapping complexity. We compare the results with other constellations found through numerical optimizations, which require significantly higher complexity than the proposed scheme.

\subsection{Bit-wise Mutual Information}

Let ${\mathcal B}_b^{(i)} \subset \mathbb{F}_2^m$ denote 
a set of binary labels with its $i$th bit specified by $b\in \mathbb{F}_2$.
Let ${\cal A}$ denote a set of $M$-ary PAM constellation points
that are mapped from the binary label through~\eqref{eq:mapping}.
The BMI of the $i$th binary channel
at a specified channel SNR $\gamma$ 
over an AWGN channel
is then expressed as~\cite{szczecinski2015bit,caire1998bit}%
\begin{align}
& C_i
\left(\mathcal{A},\gamma \right)
 = 1-\sum_{b \in \mathbb{F}_2} 
\sum_{{\bf b} \in \mathcal{B}_b^{(i)}}
P({\bf b})\,
\nonumber\\
& 
\times 
\int_{-\infty}^{\infty}
p_{Y\mid X} (y \mid \varphi({\bf b}))
\log _{2} \frac{\sum_{{\bf b}' \in \mathbb{F}_2^m}
p_{Y\mid X}(y \,|\, \varphi({\bf b}'))}
{\sum_{{\bf b}'' \in \mathcal{B}_b^{(i)}}
p_{Y\mid X}(y \,|\, \varphi({\bf b}''))} dy,
\label{eq:bitbmi}
\end{align}
where 
$P({\bf b}) = \frac{1}{M}$ (i.e., the input binary labels are equiprobable)
and 
the conditional probability density function~(PDF) 
$p_{Y|X}(y\,|\, x)$ is defined as
\begin{align}
p_{Y|X}(y\,|\, x) = \frac{1}{\sqrt{\pi N_0}} e^{-\frac{(y-x)^2}{N_0}},
\label{eq:cpdf}
\end{align}
with $N_0$ representing the power spectral density of the AWGN, which is related to $\gamma$ by 
$\gamma = E_s/N_0$.
Note that The above integral 
can be efficiently evaluated using Gaussian–Hermite 
quadrature (GHQ)~\cite{alvarado2018achievable}.
Accordingly, 
the achievable information rate of a BICM system is defined as the sum of the BMIs over all binary channels:
\begin{equation}
C
\left(\mathcal{A},\gamma \right)
 = \sum_{i=0}^{m-1}C_i(\mathcal{A}, \gamma),
\label{eq:bmi}
\end{equation}
measured in bits per dimension.

\begin{table}[tbp]
\begin{center}
    \caption{
BMI of individual binary channels for $(32,N)$-PAM 
with $\rho=1$.
The bit $b^{(0)}$ corresponds to the MSB.}
    \label{bmieb}
\begin{tabular}{|c||c|c|c|c|c|c|}
    \hline
    $N$ (SNR) & $b^{(0)}$ & $b^{(1)}$ & $b^{(2)}$ & $b^{(3)}$ & $b^{(4)}$ & Total \\ \hline \hline
    32 (10\,dB) & 0.6940 & 0.5163 & 0.2291 & 0.0741 & 0.0268 & 1.5403 \\ \hline
    10 (10\,dB) & 0.7518 & 0.5915 & 0.1739 & 0.0695 & 0.0262 & 1.6128
    \\ \hline \hline
    32 (20\,dB) & 0.9013 & 0.8453 & 0.7193 & 0.4800 & 0.2239 & 3.1699 \\ \hline
    10 (20\,dB) & 0.9990 & 0.9953 & 0.4812 & 0.2241 & 0.1158 & 2.8154
    \\ \hline 
\end{tabular}
\end{center}
\vspace{-0.5em}
\end{table}

\subsection{Effect of Constellation Point Merging on BMI}

To elucidate the validity of the proposed merging operation, we first demonstrate 
its impact on the BMI of individual binary channels.
As an example of $M=32$ (corresponding to five binary channels),
Table~\ref{bmieb} lists the BMI values of individual
binary channels for $(32,32)$-PAM and $(32,10)$-PAM 
($N=32$ and $N=10$),
each evaluated at SNRs of $10$\,dB and $20$\,dB, 
where $b^{(0)}$ and $b^{(4)}$ correspond to the MSB and the LSB, respectively.
It is observed 
in both SNR cases
that the BMIs of the lower-order bits, such as 
$b^{(3)}$ and $b^{(4)}$, 
are reduced by the merging process, while those of the higher-order bits, $b^{(0)}$ and $b^{(1)}$, are increased. 
Therefore, the proposed merging operation enhances the reliability of the higher-order bits at the expense of that of the lower-order bits.
In the case of lower SNR ($10$\,dB), in particular, 
the lower-order bits are inherently unreliable and contribute only marginally to the total BMI.
As a result, the loss in their BMIs due to merging is relatively small, whereas the enhanced separability among constellation points
successfully 
increases the BMIs of the higher-order bits, leading to an overall improvement in the total BMI.

\subsection{Loss From the Shannon Limit}

\begin{figure}[tbp]
\begin{center}
	\includegraphics[width=\hsize]{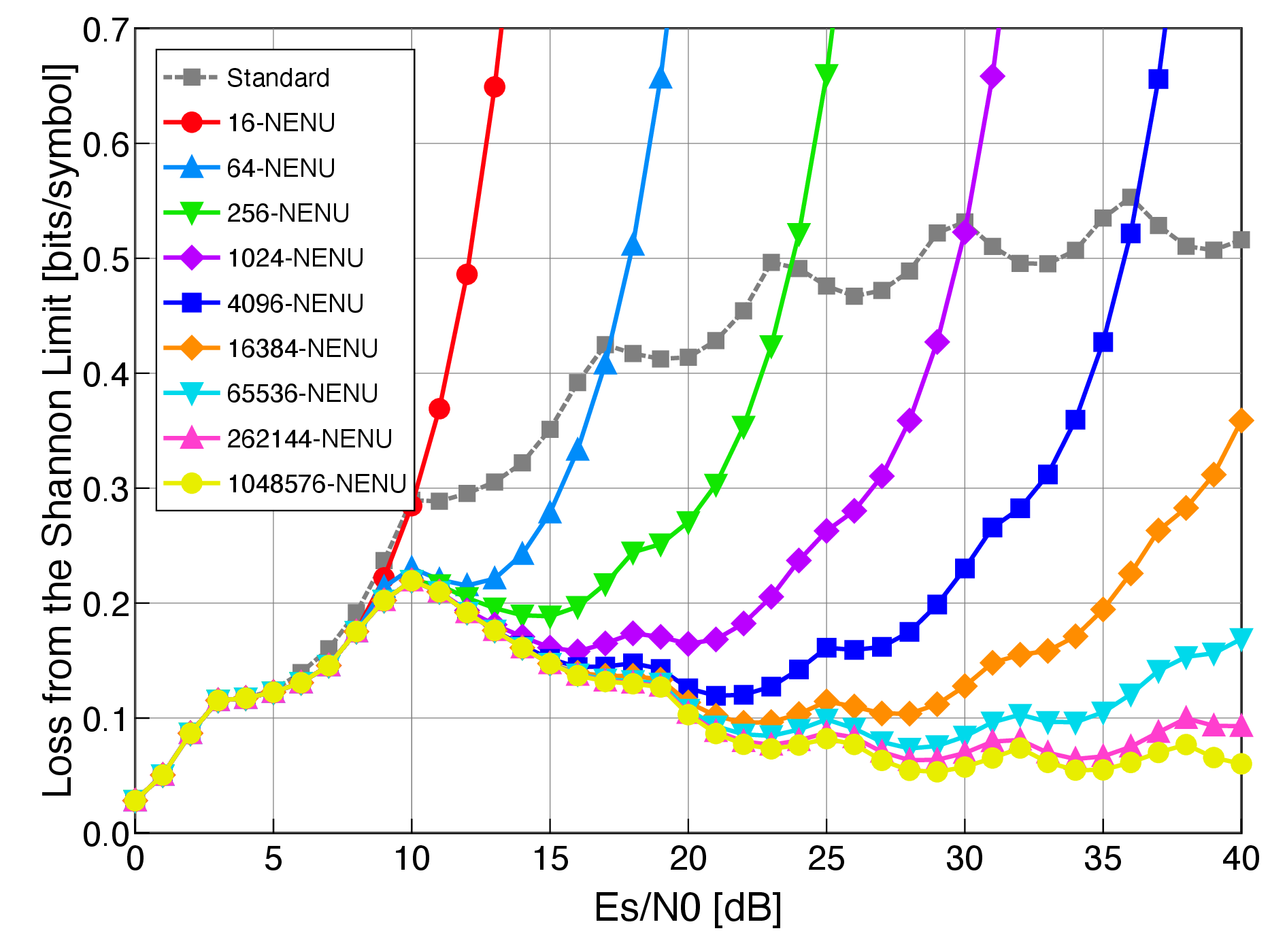}
	\caption{Loss in achievable information rate from the Shannon limit for the BICM systems using the proposed NENU constellations with various modulation orders. }
	\label{fig:GainHigh}
\end{center}
\vspace{-0.5em}
\end{figure}

We evaluate the gain achieved by  the proposed shaping scheme when its parameters are properly optimized. 
For a given pair of modulation order $M$ and operating SNR $\gamma$, the optimization problem for our NENU constellation is formulated as
\begin{equation}
\begin{aligned}
    \max_{N,\rho} \quad & 
C
\Bigl({\cal A}\left(M,N,\rho\right), \gamma \Bigr), \\
\textrm{subject to} \quad & N \in \{ 2,4,\ldots,M \} \\ & \rho \in [0,1].
\end{aligned}
\label{eq:opt}
\end{equation}
In this work, constellations with varying numbers of distinct points $N \in \{2,4,\ldots,M\}$ 
are generated for each quantized value of $\rho$, using a step size of $0.01$, in accordance with the proposed merging principle. The optimal constellation is determined by evaluating the BMI 
in~\eqref{eq:bmi}  
across all candidate parameter pairs $(N, \rho)$. 
It is important to note that if $\rho$ is fixed for a specific target SNR and $N$ is subsequently adjusted to enhance the BMI, the optimization process would be simplified. However, this approach may lead to a suboptimal constellation.

The gain achieved by constellation shaping is typically assessed  by the difference between the achievable information rate and the Shannon limit, known as the \textit{loss from the Shannon limit}~\cite{montorsi2016design}.
When the two identical $M$-ary PAM constellations form $M^2$-ary QAM,
the loss from the Shannon limit is defined 
as a function of SNR $\gamma$ by
\begin{equation}
L \left({\cal A}, \gamma\right)
\triangleq
    \operatorname{log_2}\left(1+\gamma\right) - 
2\,
C
\left({\cal A}, \gamma\right)
\label{eq:loss}
\end{equation}
in bits per symbol, where the second term corresponds to the total BMI of the two-dimensional QAM constellation. 

The loss from the Shannon limit of the proposed shaping scheme is shown in Fig.~\ref{fig:GainHigh} for various modulation orders ranging from $M^2 = 16$ to $M^2 = 1\,048\,576$.
Each plotted point is obtained by selecting the optimal parameter set $\left(N,\rho\right)$ 
at a given SNR. As a comparison, the minimum loss achieved by the uniform $M^2$-QAM 
with the optimal modulation order is indicated as a gray dashed line.
At low SNRs, the performance of the proposed shaping scheme coincides with that of conventional QAM, due to the optimality of low-order constellations such as QPSK and 
standard uniform 16-QAM in this SNR regime.
However, as the SNR increases, the gap between the BMI of standard QAM and the Shannon limit becomes increasingly pronounced.
On the other hand,
the proposed NENU constellation achieves a minimal loss from the Shannon limit, particularly in the high-SNR regimes, by exploiting higher modulation orders. 
Specifically, losses of less than $0.1$~bits per symbol are observed for modulation orders of $16\,384$ and above.
These remarkable gains are attributed to the high-resolution Gaussian approximation enabled by the huge number of constellation points.
Moreover, higher-order NENU constellations consistently outperform lower-order ones across the entire SNR range.
This confirms that the proposed merging scheme adaptively reduces the effective constellation size of high-order NENU constellations in the low-SNR regime.

\subsection{Comparison With Other Geometric Shaping Schemes} 

\begin{figure}[tbp]
\begin{center}
	\includegraphics[width=\hsize]{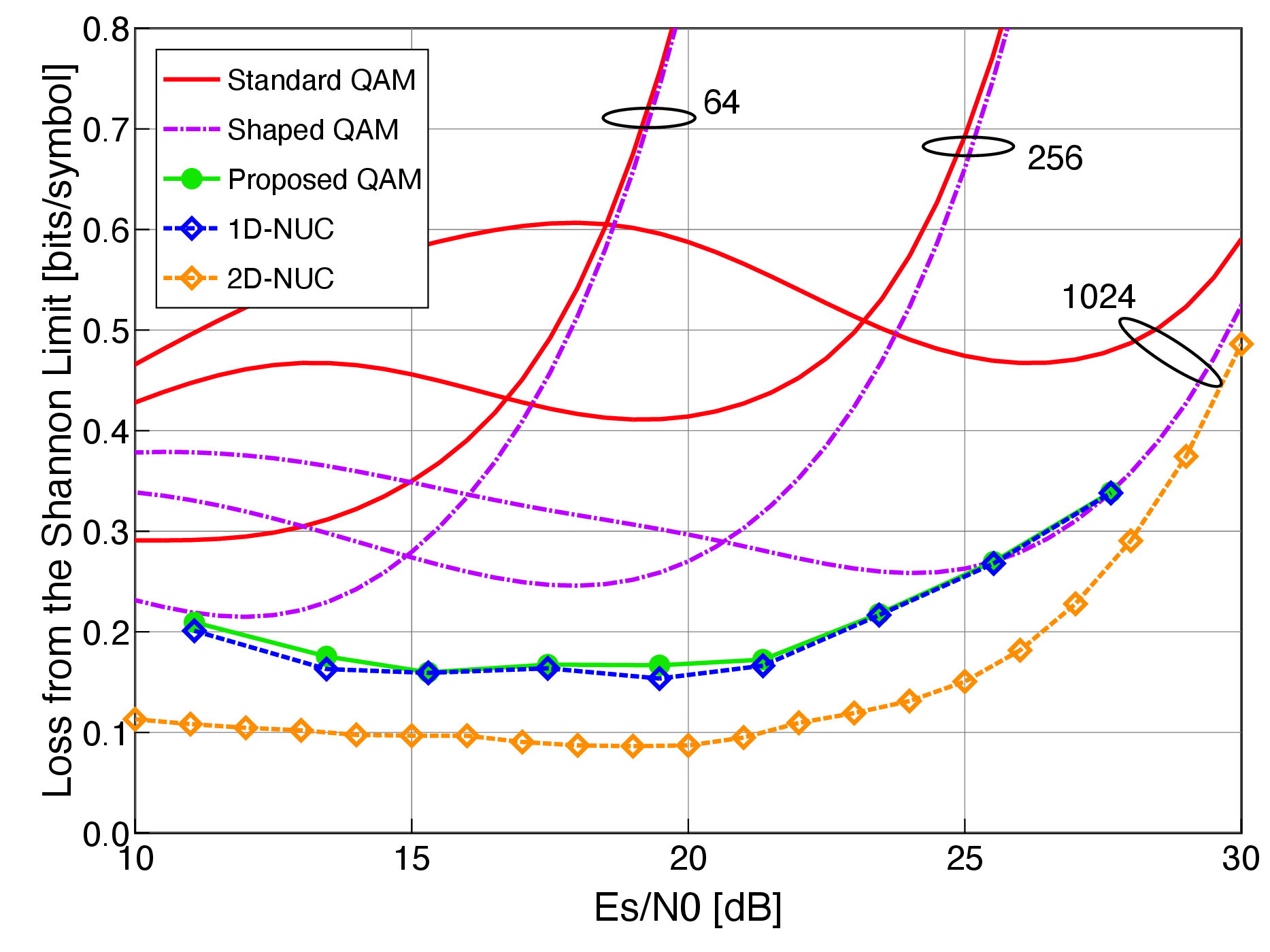}
	\caption{Loss from the Shannon limit for the BICM systems with several different modulation schemes.}
	\label{fig:Gain}
\end{center}
\vspace{-0.5em}
\end{figure}

We next compare the proposed NENU constellations with other representative modulation schemes.
The following 
constellations are considered for comparison:
\begin{itemize}
\item The uniform $M^2$-ary QAM with $M^2 \in \{ 64, 256,1024\}$
\item The proposed one-dimensional shaping without constellation point merging, i.e., ${\cal A}\left(M,M,\rho\right)$ with $M^2 \in \{ 64, 256,1024\}$~(indicated as {\em shaped QAM}).
\item The proposed NENU constellation ${\cal A}\left(32,N,\rho\right)$.
\item The one-dimensional~(1D) $1024$-NUC of ATSC~3.0.
\item The two-dimensional~(2D) $1024$-NUC developed in~\cite{sillekens2022high}.
\end{itemize}
The corresponding loss from the Shannon limit is compared 
in Fig.~\ref{fig:Gain}, where
all shaping schemes 
other than standard uniform QAM
are optimized at each plotted SNR according to their respective design or optimization algorithms.

We begin with a brief discussion of
the design complexity 
in terms of 
the number of 
real-valued variables to be optimized (i.e., the degrees of freedom). 
Let $m' = 2m = \log_2 M^2$ denote the number of binary channels 
(i.e., the maximum spectral efficiency in bits per symbol).
For the one-dimensional NUC in ATSC, $2^{m'/2-1} = 2^{m-1}$ real-valued PAM amplitudes 
need to be
optimized under an origin-symmetry constraint.
On the other hand, the two-dimensional NUC in~\cite{sillekens2022high} is designed directly in the complex plane under a quadrant symmetry. 
In this case, $2^{m'-2} = 2^{2(m -1)}$ 
complex-valued constellation points in a single quadrant 
must be 
optimized, which corresponds to $2^{2m-1}$ real-valued variables for the I/Q components.
In contrast, for the proposed NENU constellation, all PAM amplitudes are optimized using only two real-valued parameters irrespective of $m$, highlighting its scalability for high-spectral-efficiency regimes.

As shown in Fig.~\ref{fig:Gain},
the analysis of the achievable rate reveals
that the optimal modulation order in BICM is influenced by the channel SNR.
The implementation of shaped QAM without merging constellation points has been validated to provide better performance compared to standard uniform QAM~\cite{kurihara2024design}. 
However, the limitation of constellation sizes being restricted to powers of two results in notable gaps from the Shannon limit at certain SNR ranges, specifically around 15\,dB and 20\,dB.
In contrast, the proposed shaping scheme maintains these gaps at a minimal level over a broad 
range of SNRs. 
This indicates that using shaping with the ability to flexibly adjust both the constellation size and its distribution is beneficial for a rate-adaptive BICM design. 
The performance gain from the proposed shaping scheme is comparable to 
that of the one-dimensional NUC in ATSC~3.0 across all SNR ranges, despite the latter being achieved through a more complex optimization process for a specific target SNR or information rate.

When comparing the results with the two-dimensional constellation of~\cite{sillekens2022high},
performance gaps of approximately $0.08$ bits per symbol are observed, which highlights the inherent limitations of one-dimensional constellation design.
However, 
one-dimensional modulation schemes offer practical advantages in very high data-rate regimes,
since the demapping process for two-dimensional constellations requires a substantial 
increase in computational cost with higher modulation orders, as is discussed in the next subsection.

\subsection{Demapping Complexity}
\label{sec:complexity}

In BICM systems, soft demapping is carried out by computing the bit log-likelihood ratios (LLRs) for each binary channel, which dominates the computational complexity at the receiver.
The required computational effort strongly depends on the geometric structure of the underlying constellation and becomes increasingly critical as the modulation order increases.
Accordingly, we analyze the LLR computation complexity for general one-dimensional and two-dimensional constellations and then discuss the corresponding complexity of the proposed NENU constellation.

Let us assume that the transmitted symbols are selected from a set of specific complex-valued 
(i.e., two-dimensional) constellation 
$\mathcal{X}$, 
with the number of points represented by 
$|\mathcal{X}|$.
The LLR value for the $i$th binary channel is computed as 
\begin{align}
L_i &= \log \left[\frac{\sum_{x \in \mathcal{X}_{0}^{(i)}} p(x)\exp \left(-\frac{|y-x|^2}{N_0}\right)}{\sum_{x \in \mathcal{X}_{1}^{(i)}} p(x) \exp \left(-\frac{|y-x|^2}{N_0}\right)}\right] \nonumber \\
&\approx \frac{1}{N_0}\left( \min _{x \in \mathcal{X}_1^{\left(i\right)}}\left\{|y-x|^2\right\}-\min _{x \in \mathcal{X}_0^{\left(i\right)}}\left\{|y-x|^2\right\} \right),
\label{eq:Lmin}
\end{align}
where $y \in {\mathbb{C}}$ denotes the received complex-valued symbol
and $\mathcal{X}_{b}^{(i)}
\subset \mathcal{X}$ represents the set of
constellation points in the $i$th binary channel with the bit specified by $b \in \mathbb{F}_2$.
On the other hand, for a one-dimensional constellation, where $\mathcal{X}$ is composed of two identical and orthogonal PAM constellations $\mathcal{A}$,
the LLR value can be expressed as
\begin{align}
L_i &\approx \frac{1}{N_0}\left( \min _{x \in \mathcal{A}_1^{\left(i\right)}}\left\{(y-x)^2\right\}-\min _{x \in \mathcal{A}_0^{\left(i\right)}}\left\{(y-x)^2\right\} \right),
\label{eq:Lmin2}
\end{align}
where $y \in {\mathbb{R}}$ is the received real-valued symbol,
$\mathcal{A}_{b}^{(i)}
\subset \mathcal{A}$ denotes the set of
constellation points in the $i$th binary channel with the bit specified by $b \in \mathbb{F}_2$,
and $x$ is thus chosen from a one-dimensional (i.e., real-valued) constellation.

We focus only on the number of required multiplication operations, since the complexity associated with comparison and addition operations is negligible compared to that of multiplication operations.
In the case of a two-dimensional constellation,
from~\eqref{eq:Lmin} it is necessary 
to compute the 
squared Euclidean distances
between the received symbol and 
each of the $|\mathcal{X}|=M^2$ candidate constellation points. 
This computation entails performing $2|\mathcal{X}|$ real multiplications.
Meanwhile, in the case of a one-dimensional constellation,
calculating~\eqref{eq:Lmin2} requires
$|\mathcal{A}|$ real multiplications.
The square root of the cardinality of a two-dimensional constellation, 
$|\mathcal{A}|=M$, represents the corresponding cardinality in this case. 
Therefore, as $M$ increases to support higher spectral efficiency, the use of a one-dimensional constellation should be critical from an implementation complexity perspective.
This reasoning may elucidate the rationale behind 
the adoption of one-dimensional constellations for high-order scenarios in practical wireless communication systems such as ATSC~3.0. 

\begin{figure}[tbp]
\begin{center}
	\includegraphics[width=\hsize]{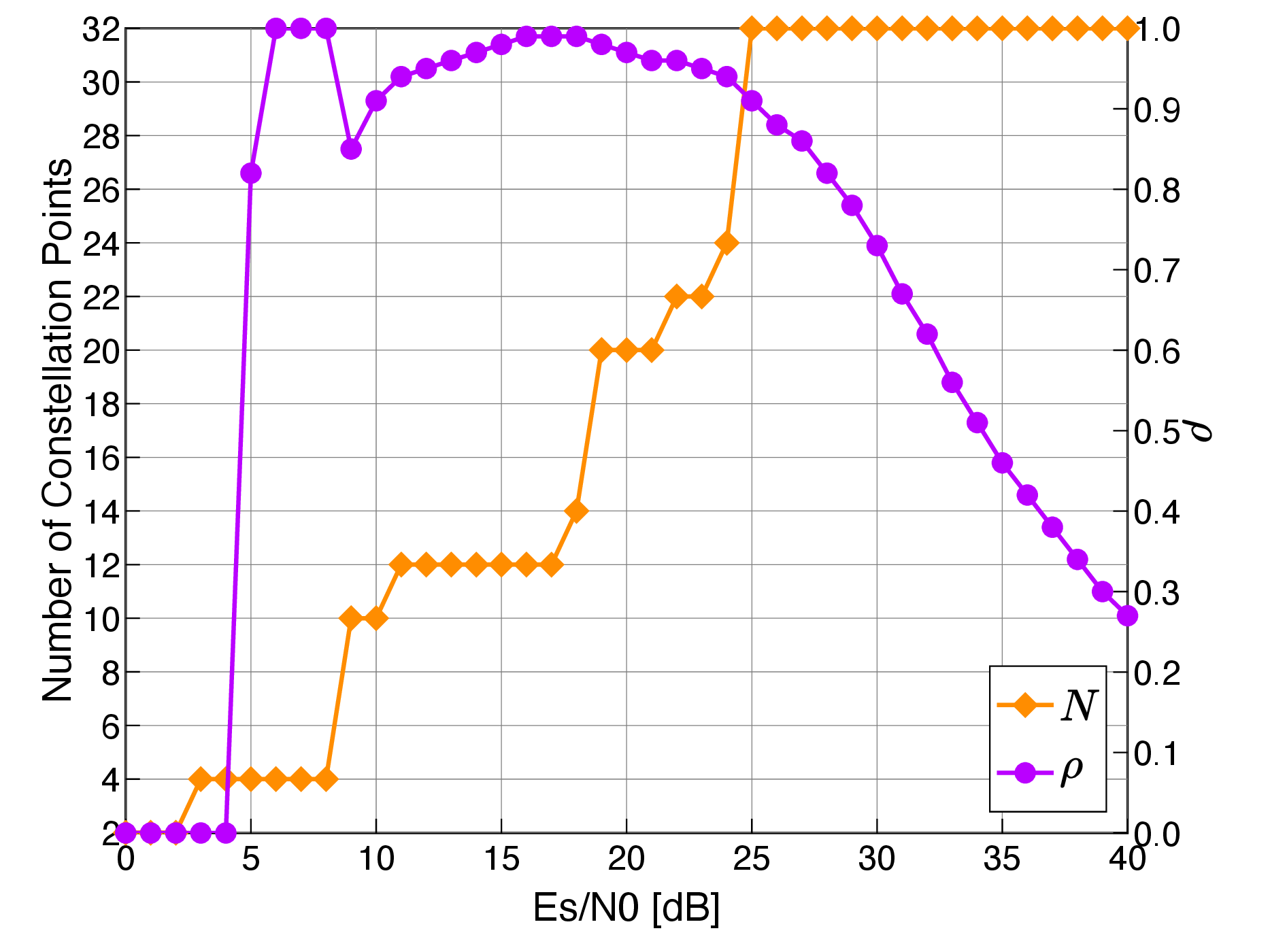}
	\caption{The optimal values of constellation size $N$ and design parameter $\rho$ according to the channel SNR ($M=32$).}
	\label{fig:NRCP}
\end{center}
\vspace{-0.5em}
\end{figure}

The proposed NENU constellation also offers a more efficient demapping implementation due to 
its reduced constellation cardinality. 
In the LLR computation in~\eqref{eq:Lmin2}, since Euclidean distances are compared among $|\mathcal{A}(M,N,\rho)|=N$ candidate constellation points, 
the number of required real multiplications is given by $N$.
With the modulation order set to $M=32$, the optimal constellation size $N$ and the design parameter $\rho$ as a function of channel SNR $\gamma$ are illustrated in Fig.~\ref{fig:NRCP}.
It is observed that the required number of distinct signal points reduces as the SNR decreases.
It thus demonstrates that the proposed NENU constellation not only achieves near-optimal gain but also minimizes the demapping complexity.

We note that several simplified demapping algorithms have been developed for both one-dimensional and two-dimensional NUCs~\cite{fuentes2015low,hong2020backward,hong2022enhanced,barrueco2017condensation,duan2021non,barrueco2018low}.
Quadrant search reduction (QSR)~\cite{fuentes2015low} reduces the number of the required Euclidean distance computations by discarding many symbols in different quadrants that 
have negligible impact on the LLR.
Virtual point searching (VPS)~\cite{hong2020backward,hong2022enhanced}, on the other hand, approximates a general two-dimensional constellation by 
using virtual points located on several radial axes, similar to the
amplitude-phase shift keying (APSK) constellation.
While this approximation can significantly reduce the demapping complexity, 
it 
could lead 
to a substantial degradation in the achievable information rate.
Condensed symbol reduction (CSR)~\cite{barrueco2017condensation,duan2021non,barrueco2018low} reduces the demapping complexity by exploiting the clustering of constellation points that typically appears in 
NUCs optimized for 
low code rates, 
where each cluster is treated as a single 
symbol in the LLR computation.
CSR can be applied to both one-dimensional and two-dimensional NUCs.
In the case of two-dimensional NUCs, it is sometimes combined with VPS or QSR to further reduce the complexity~\cite{fuentes2015low,hong2020backward,hong2022enhanced}.

Nevertheless, the most effective reduction in demapping complexity is still achieved by exploiting
I/Q independence, which naturally favors one-dimensional constellations especially for high modulation orders.
In this context, 
a one-dimensional NUC with
a carefully designed CSR-based demapper
may achieve performance comparable to the proposed NENU constellation.
However, it should rely on heuristic procedures 
that determine the initial constellation points and perform the clustering process.
In contrast, 
since
the proposed NENU constellation is uniquely described by a simple mathematical expression of~\eqref{eq:mapper}, 
it ensures systematic reproducibility and enables a significantly simpler optimization process.

\section{Application to BICM-OFDM With CAF and CNC}
\label{sec:System}

The design of wireless signals with low 
PAPR continues to be a prominent and challenging field of research~\cite{9725261,9747919,9962325,10530524}.
BICM-OFDM is a {\em de facto} standard for high data rate wireless communications
over frequency-selective fading channels, but due to its multicarrier nature,
its signal has high PAPR regardless of the constellations employed.
Thus, PAPR reduction should be necessary for better power amplifier efficiency.
Among many PAPR reduction approaches developed for OFDM, we suggest employing 
CAF~\cite{ochiai2002performance}, as it is directly applicable to the 
BICM-OFDM system using the proposed 
NENU constellation.
The major drawback is its nonlinear distortion, which should be compensated for at the receiver.
Therefore, 
we address the use of CNC~\cite{chen2003iterative,sun2021},
which is also compatible with our constellation shaping. 
We demonstrate that the proposed NENU constellations can be flexibly optimized 
for such systems, achieving notable gains under practical scenarios.

In what follows, for simplicity of notation, we define the set of QAM symbols composed 
of the two NENU PAM constellations as
\begin{equation}
{\cal X}_{M,N,\rho} \triangleq \{
A_I + j A_Q \,|\, A_I, A_Q \in {\cal A}\left(M,N,\rho\right)
\}.
\nonumber
\end{equation}

\subsection{OFDM Transmitter With CAF}

\begin{figure}[tbp]
\begin{center}
	\includegraphics[width=\hsize]{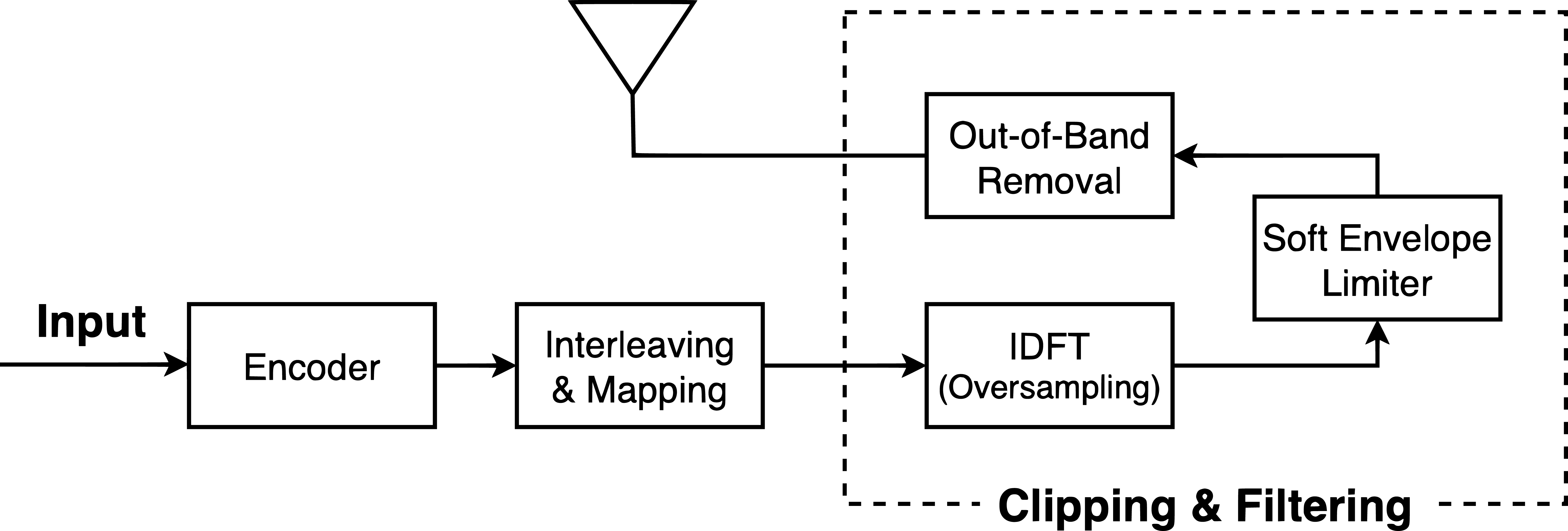}
    \caption{Block diagram of the BICM-OFDM transmitter with CAF.
CAF is implemented through a sequence of operations: oversampling using a $JN_c$-point IDFT, soft envelope limiter, and out-of-band signal removal.}
    \label{fig:modelCAF}
\end{center}    
\vspace{-0.5em}
\end{figure}

We follow the description of~\cite{ochiai2002performance} for CAF and
that of~\cite{sun2021} for CNC
with an extension to coded OFDM.
The block diagram of the OFDM transmitter with CAF is illustrated in Fig.~\ref{fig:modelCAF}.
The information sequence is first encoded by a binary channel code with 
code rate $R_c \in (0,1]$. 
The codeword is bit-wise interleaved and 
then mapped onto the corresponding NENU constellations.
Let 
$\mathbf{X}=\left(X_{0}, X_{1}, \ldots, X_{N_{c}-1}\right) \in \mathcal{X}_{M,N,\rho}^{N_c}$ 
denote a vector of 
QAM symbols carried by each $N_{c}$-subcarrier OFDM signal, where 
$X_{k} = A_{I,k} + j A_{Q,k}$
with $A_{I,k}$ and $A_{Q,k}$ representing NENU PAM symbol on the $k$th subcarrier.
The $J$-times oversampled OFDM symbol $\mathbf{S} = (S_0,S_1,...,S_{JN_{c}-1})
\in {\mathbb C}^{J N_c}$ is generated 
by performing $JN_{c}$-point inverse discrete Fourier transform (IDFT) on 
the zero-padded symbol vector $\mathbf{X}'=(\mathbf{X},
\underbrace{
0,0,\ldots,0}_{(J-1)N_c}) \in {\mathbb C}^{J N_c}$
as follows:\footnote{
In practice, DFT and IDFT are implemented using the fast Fourier transform~(FFT) algorithm
by setting the number of points $J N_c$ as the powers of two.
}
\begin{equation}
S_{n}=\frac{1}{\sqrt{J N_{c}}} \sum_{k=0}^{J N_{c}-1} X'_{k} e^{j 2 \pi k \frac{n}{J N_{c}}}, \quad n=0,1, \ldots, J N_{c}-1 .
\end{equation}
To restrict the peak power, 
these time domain signal samples are clipped by soft-envelope limiter~\cite{ochiai2002performance}
as
\begin{equation}
\tilde{S}_{n}= \begin{cases}S_{n}, & \text { for }\left|S_{n}\right| \leq r_\text{max}, 
\\ r_\text{max} e^{j \arg S_{n}}, & \text { for }\left|S_{n}\right|>r_\text{max},
\end{cases}
\end{equation}
where $r_\text{max}$ represents 
the maximum envelope level. We define
the clipping ratio $\gamma_\text{CR}$ as
\begin{equation}
\gamma_\text{CR} \triangleq \frac{r_\text{max}}{\sqrt{P_{\text {in}}}} ,
\end{equation}
where 
$P_\mathrm{in} \triangleq \mathbb{E}\left[\|\mathbf{X}\|^{2}\right]/N_{c}$ 
is the average power of the OFDM signal prior to CAF.
We then perform $JN_{c}$-point discrete Fourier transform (DFT) on the clipped signals 
to transform them back to 
the frequency-domain symbols represented by 
$\mathbf{U}^{\prime}=\left(
U'_{0}, U'_{1}, \ldots, U'_{J N_{c}-1} \right) \in {\mathbb C}^{J N_c}$,  
where
\begin{equation}
U'_k=\frac{1}{\sqrt{J N_{c}}} \sum_{n=0}^{J N_{c}-1} \tilde{S}_{n} e^{-j 2 \pi k \frac{n}{J N_{c}}},
\,\, k=0,1, \ldots, J N_{c}-1.
\end{equation}
By removing the out-of-band component from these symbols, we obtain the in-band component 
${\mathbf{U}}=\left(
{U}_{0}, {U}_{1}, \ldots, {U}_{N_{c}-1}\right)
\in {\mathbb C}^{N_c}$ with
${U}_{k} \equiv {U}'_{k}$
as the transmit symbols.

The average power of the transmit OFDM signal after 
the CAF process is given by 
$P_\mathrm{av} \triangleq \mathbb{E}\left[\|\mathbf{{U}}\|^{2}\right] /N_{c}$,
i.e., the average signal power that includes the in-band distortion caused by clipping.
For a fair comparison in terms of the transmit signal power and the channel noise variance,
we define the SNR of an AWGN channel as 
\begin{equation}
\mathrm{SNR} \triangleq \frac{P_{\mathrm{av}}}{N_0} .
\end{equation}
The above definition of SNR ensures that 
the signal part is defined as the actual transmit signal including the in-band distortion (clipping noise).

\subsection{OFDM Receiver With CNC}

\begin{figure}[tbp]
\begin{center}
	\includegraphics[width=\hsize]{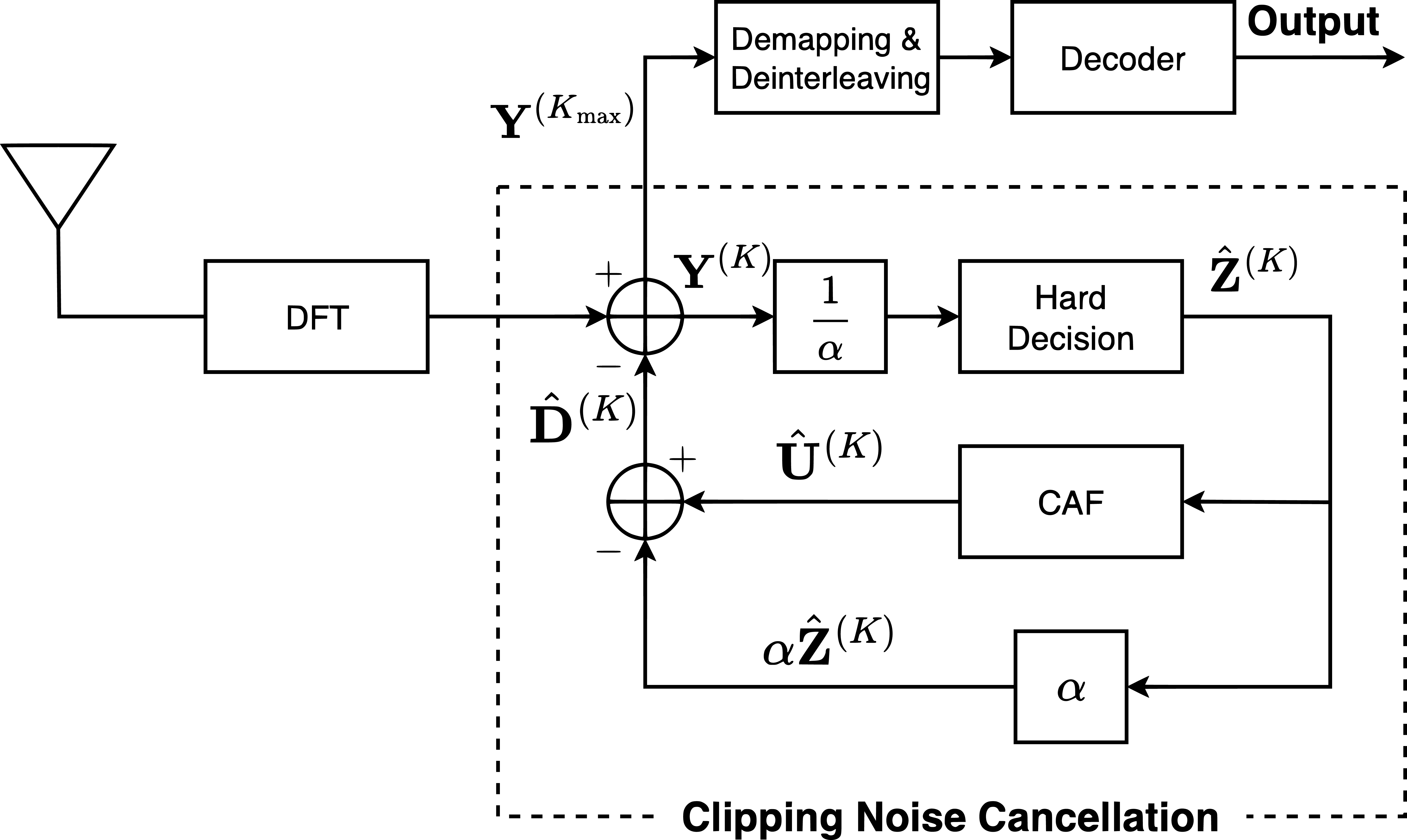}
    \caption{Block diagram of the BICM-OFDM receiver with CNC.
CNC iteratively estimates the clipping noise based on the regenerated clipped OFDM signal. Each iteration involves estimating the transmitted symbols and applying the CAF process, including $J N_c$-point IDFT and DFT. }
    \label{fig:modelCNC}
\end{center}    
\vspace{-0.5em}
\end{figure}

At the receiver, we consider performing CNC 
to mitigate the in-band distortion caused by CAF prior to channel decoding\footnote{%
It is possible to design CNC that takes into account the channel decoder output
as well as re-encoder in its loop, but this approach makes not only the receiver complexity
and latency prohibitive, but also the appropriate system design more challenging.
}.
The block diagram of the OFDM receiver with CNC is illustrated in Fig.~\ref{fig:modelCNC}.
In the case of an AWGN channel, the received symbol corresponding to the $k$th 
subcarrier can be expressed as
\begin{equation}
\tilde{Y}_k =\alpha X_k  + D_k+ W_k, 
\end{equation}
where $D_k \in {\mathbb C}$ 
denotes the distortion component caused by the CAF process 
and $W_k \in {\mathbb C}$
represents the Gaussian noise with $W_k \sim \mathcal{CN}\left(0, N_0\right)$, 
both corresponding to the $k$th subcarrier. 
Assuming that the baseband OFDM signal follows a complex Gaussian distribution, 
the attenuation factor $\alpha$ due to clipping 
can be expressed as~\cite{ochiai2002performance}
\begin{equation}
\alpha=1-e^{-\gamma_\text{CR}^{2}}+\frac{\sqrt{\pi} \gamma_\text{CR}}{2} \operatorname{erfc}(\gamma_\text{CR}),
\end{equation}
where $\operatorname{erfc}(x) =1 -\operatorname{erf}(x)$.
Provided that the length of cyclic prefix is sufficient compared
to that of the channel impulse response,
the received symbol over a frequency-selective fading channel
can be expressed as
\begin{equation}
Y_k =
H_k \left(\alpha X_k  + D_k \right) + W_k, 
\end{equation}
where $H_k$ is the channel coefficient of the $k$th subcarrier.
Assuming that the channel coefficients can be perfectly estimated at the receiver, 
we may apply the zero-forcing (ZF) equalization as
\begin{equation}
\tilde{Y}_k \triangleq 
\frac {Y_k}{H_k} = \alpha X_k + D_k  + W_k^{\prime}, 
\end{equation}
where $W_k^{\prime} \triangleq  W_k/H_k \sim 
\mathcal{CN}\left(0, N_0/\left|H_{k}\right|^{2} \right)$.

In the CNC process, we iteratively estimate and remove the 
distortion component 
${\mathbf{D}}=\left(
{D}_{0}, {D}_{1}, \ldots, {D}_{N_{c}-1}\right)\in {\mathbb C}^{N_c}$. 
The CNC algorithm at the $K$th iteration (starting with $K:=0$) is described as follows:
\begin{enumerate}
    \item Apply the hard decision based on the constellation set $\mathcal{A}({M,N,\rho})$ 
to the equalized and normalized received signal $Y_k^{(K)}/\alpha$ (with $Y_k^{(0)}:=\tilde{Y}_k$) to estimate the transmitted symbol $
\hat{X}^{(K)}_{k}={\hat{A}_{I,k}^{(K)}}+j{\hat{A}_{Q,k}^{(K)}}$, which is given by
    \begin{align}
\begin{cases}
{\hat{A}}_{I,k}^{(K)} =\arg \min _{{A} \in \mathcal{A}({M,N,\rho})}\left(
{A}- \Re\left\{
{{Y}}_k^{(K)} \right\}/ \alpha \right)^{2}, \\
{\hat{A}}_{Q,k}^{(K)}  =\arg \min _{{A} \in \mathcal{A}({M,N,\rho})}\left(
{A}- \Im\left\{{{Y}}_k^{(K)} \right\}/ \alpha \right)^{2} .
\end{cases}
\label{eq:hard}
\end{align}
    \item  Apply the same CAF process at the transmitter to obtain the estimated in-band component, which is modeled as 
    \begin{equation}
    \hat{{U}}_{k}^{(K)}=\alpha \hat{{X}}_{k}^{(K)}+\hat{{D}}_{k}^{(K)} .
    \label{eq:CAFinCNC}
    \end{equation}

    \item Calculate the distortion component by
    \begin{equation}
\hat{{D}}_{k}^{(K)}=\hat{{U}}_{k}^{(K)}-\alpha \hat{{X}}_{k}^{(K)},
\end{equation}
    and subtract it from the original received symbol as
    \begin{align}
{{Y}}_{k}^{(K+1)} & =\tilde{Y}_{k}-\hat{{D}}_{k}^{(K)} \nonumber\\
& =\alpha {X}_{k}+\left({D}_{k}-\hat{{D}}_{k}^{(K)}\right)+{W}_{k} .
\end{align}
    
    \item Set $K:=K + 1$.
If $K$ reaches the iteration limit $K_\text{max}$, output ${{Y}}_{k}^{(K_\text{max})}$ and stop. Otherwise, go to Step 1). 
\end{enumerate}
Note that each iteration involves CAF at the receiver side, which requires 
performing $J N_c$-point IDFT, clipping, and $J N_c$-point DFT at Step 2).

If the majority of the estimated symbols are correct, the residual distortion component ${D}_{k}-\hat{{D}}_{k}^{(K)}$ should diminish as $K$ increases. This cancellation process may be successful
when the channel SNR is sufficiently high.
Since our target is a high spectral efficiency regime, this high-SNR assumption naturally applies,
and thus CNC is expected to work properly, provided that the clipping ratio $\gamma_\text{CR}$ 
is appropriately selected. 
Although a lower clipping ratio helps reduce PAPR and improve power amplifier efficiency, it introduces more severe nonlinear distortion, which may prevent CNC from functioning properly.

Note that since our NENU constellations are designed based on a one-dimensional constellation, the complexity associated with the hard decision in \eqref{eq:hard} 
is
considerably
lower than that of a two-dimensional constellation, similar to the argument of demapper 
complexity discussed in Section~\ref{sec:complexity}. 
In fact, if the modulation scheme is based on a general two-dimensional constellation denoted by $\mathcal{X}$,
the hard decision must be performed in the complex plane as
\begin{equation}
{\hat{X}}_{k}^{(K)} =\arg \min _{{X} \in \mathcal{X}}\left|
{X}-
{{Y}}_k^{(K)} / \alpha \right|^{2},
\label{eq:hard2D}
\end{equation}
which requires comparing Euclidean distances among $\left|{\cal X} \right|=M^2$ symbols,
instead of $\left|
\mathcal{A}({M,N,\rho})\right|= N$ symbols.
This observation also justifies the use of one-dimensional constellations for our BICM-OFDM systems employing CAF and CNC.

\subsection{Parameter Optimization for OFDM with CAF and CNC}

The performance of OFDM with CAF and CNC 
significantly depends on system parameters, including the number of subcarriers, clipping ratio, the number of iterations, and the channel models~\cite{sun2021}.
Therefore, it can only be evaluated by 
simulations for a given system setup.
In what follows, we will first define the realistic system setup 
and then attempt to optimize shaping parameters based on computer simulations.

\subsubsection{System Setup}

The specific system parameters as well as the channel model considered through 
the remainder of this section
are listed in Table~\ref{tab:sim_setup_opt}:
We consider a specific OFDM system with
the number of subcarriers set to $N_c = 1024$, with 
each subcarrier modulated by $32$-PAM per dimension, corresponding to $1024$-QAM per subcarrier.
Channel coding is performed using an LDPC code 
compliant with
the 5G New Radio (NR) standard, with a codeword length of $10240$ bits~\cite{etsi38138}.
The code rate is fixed to $R_c = 4/5$,
which results in a spectral efficiency of $4.0$ bits per dimension, 
or $8.0$ bits per complex channel use. At the receiver, decoding is carried out using
an offset min-sum decoding algorithm.
Both AWGN and fading channels are considered.
For the fading scenario, we adopt a frequency-selective Rayleigh fading channel following the 5G NR Tapped Delay Line (TDL)-B 
model~\cite{etsi2020138}.

In the case of CAF and CNC operations, 
the oversampling factor is fixed to $J = 4$ throughout this work.
Based on the preliminary simulation with the above-mentioned system setup, 
we found that the clipping ratio of $\gamma_\text{CR} = 1.5$ 
is the minimum acceptable value that does not cause 
the CNC operation to malfunction\footnote{%
Optimal clipping ratio may be mathematically 
analyzed in terms of signal-to-noise plus distortion 
power ratio~(SNDR) when CAF is applied~\cite{azolini}, 
but since our system makes use of CNC at the receiver, its error rate performance depends
on several parameters such as the number of subcarriers and constellations~\cite{sun2021}.
As a result, it may become mathematically intractable.
}.

\begin{table}[tbp]
\begin{center}
    \caption{System parameters and channel model.}
    \label{tab:sim_setup_opt}
\begin{tabular}{|c|c|}
    \hline
    The number of subcarriers & $N_c=1024$ \\ \hline
    Modulation & $1024$-QAM ($32$-PAM each) \\ \hline
    Clipping ratio & $\gamma_\text{CR}=1.5$\\ \hline
    Oversampling factor& $J=4$\\ \hline
    Maximum CNC iterations & $K_\text{max}=10$ \\ \hline
    Channel code & 5G NR LDPC~\cite{etsi38138} \\ \hline
    Decoder & Offset min-sum decoding \\ \hline
    Code rate & $4/5$ \\ \hline
    Codeword length & 10240 \\ \hline
    Spectral efficiency & 4.0 per dimension \\ \hline
    Fading channel & 5G NR Tapped Delay Line (TDL)-B~\cite{etsi2020138} \\ \hline
\end{tabular}
\end{center}
\vspace{-0.5em}
\end{table}

\subsubsection{Parameter Optimization}

\begin{table}[tbp]
\begin{center}
    \caption{System description and optimal parameters.}
    \label{tab:modes}
\begin{tabular}{|c|c|c|c|}
    \hline
    Mode & CAF/CNC & Target Channel & Parameters $(N,\rho)$ \\ \hline \hline
    {\sf A} & not applied & AWGN/fading & $(32,0.88)$ \\ \hline
    {\sf B} & applied & AWGN & $(24,0.96)$  \\ \hline
    {\sf C} & applied & fading & $(26,0.90)$  \\ \hline
\end{tabular}
\end{center}
\vspace{-0.5em}
\end{table}

We optimize the NENU constellation for 
specific 
system configurations and channel conditions.
We consider the following three system scenarios:
\begin{itemize}
\item[{\sf A}.] Without CAF, the system operates over either an AWGN channel or a fading channel.
\item[{\sf B}.] With CAF and CNC, the system operates over an AWGN channel.
\item[{\sf C}.] With CAF and CNC, the system operates over a fading channel.
\end{itemize}
In systems that utilize CAF with $\gamma_\text{CR} = 1.5$, 
CNC is unable to completely eliminate clipping noise. 
Consequently, the effective channel becomes a mixture of AWGN and residual clipping noise, which may not possess a mathematically tractable representation. 
In subsequent studies focused on BMI evaluation and LLR computation, the noise variance $N_0$ 
specifically represents only the AWGN components. 
This limitation occurs because the residual clipping noise is affected by the constellation geometry and the transmitted symbol sequence, making it difficult to consistently incorporate it into 
$N_0$ across different modulation schemes.

Both the number of constellation points and their  distribution should be designed to match the corresponding system condition.
To this end, we evaluate the BER over a family of NENU constellations parameterized by $(N,\rho)$
and select the constellation that achieves a BER level of $10^{-4}$ at the lowest SNR as the representative design\footnote{
As discussed in Section~III, the two parameters can also be selected based on the BMI.
However, when the codeword length is not sufficient, and thus the gap between the channel capacity and the practical decoding waterfall region becomes non-negligible, 
BER-based optimization provides a more accurate criterion for parameter selection.}.
For each system scenario, the optimal parameter set $(N,\rho)$ obtained
through preliminary simulations, as well as their descriptions, are
listed in Table~\ref{tab:modes}. 
In what follows, we refer to the resulting NENU constellations corresponding 
to the above systems {\sf A}, {\sf B}, and {\sf C}
as Modes {\sf A}, {\sf B}, and {\sf C}, respectively,
and denote them by ${\mathcal{X}_{\sf A}}$, ${\mathcal{X}_{\sf B}}$, and ${\mathcal{X}_{\sf C}}$.
Here, we note that in the absence of CAF, the optimal parameter set $(N,\rho)$ 
turned out to be identical for both AWGN and fading channels, and thus they are unified
as Mode {\sf A}.
From the above results, we observe that the optimal number of constellation points depends on the channel conditions, which agrees with the observation in Section~\ref{sec:performance}.
Without CAF, $N = M=32$ is selected as optimal due to the high spectral efficiency 
setup with $R_c = 4/5$.
In contrast, with CAF and CNC, smaller values of $N$ are preferred to better accommodate the residual nonlinear distortion that reduces the effective SNR, but the optimal values may differ 
depending on the channel models.

It is worth mentioning that constellation design based on exhaustive numerical optimization algorithms, such as particle swarm optimization (PSO), does not rely on a predefined set of constellation candidates. Therefore, the optimized constellation, if found, should achieve
near-optimal performance.
However, if the objective function for optimization itself is computationally demanding,
e.g., the BER after CAF and CNC processing, the exhaustive search should become challenging.
On the other hand, due to limited degrees of freedom,
the proposed shaping scheme requires evaluation over a tractable number of constellation candidates.
It thus significantly reduces the computational burden and enables efficient optimization, 
even if the objective function is mathematically intractable and thus requires exhaustive 
simulations.

\subsubsection{Total BMI for AWGN Channel}

\begin{figure}[tbp]
\begin{center}
	\includegraphics[width=\hsize]{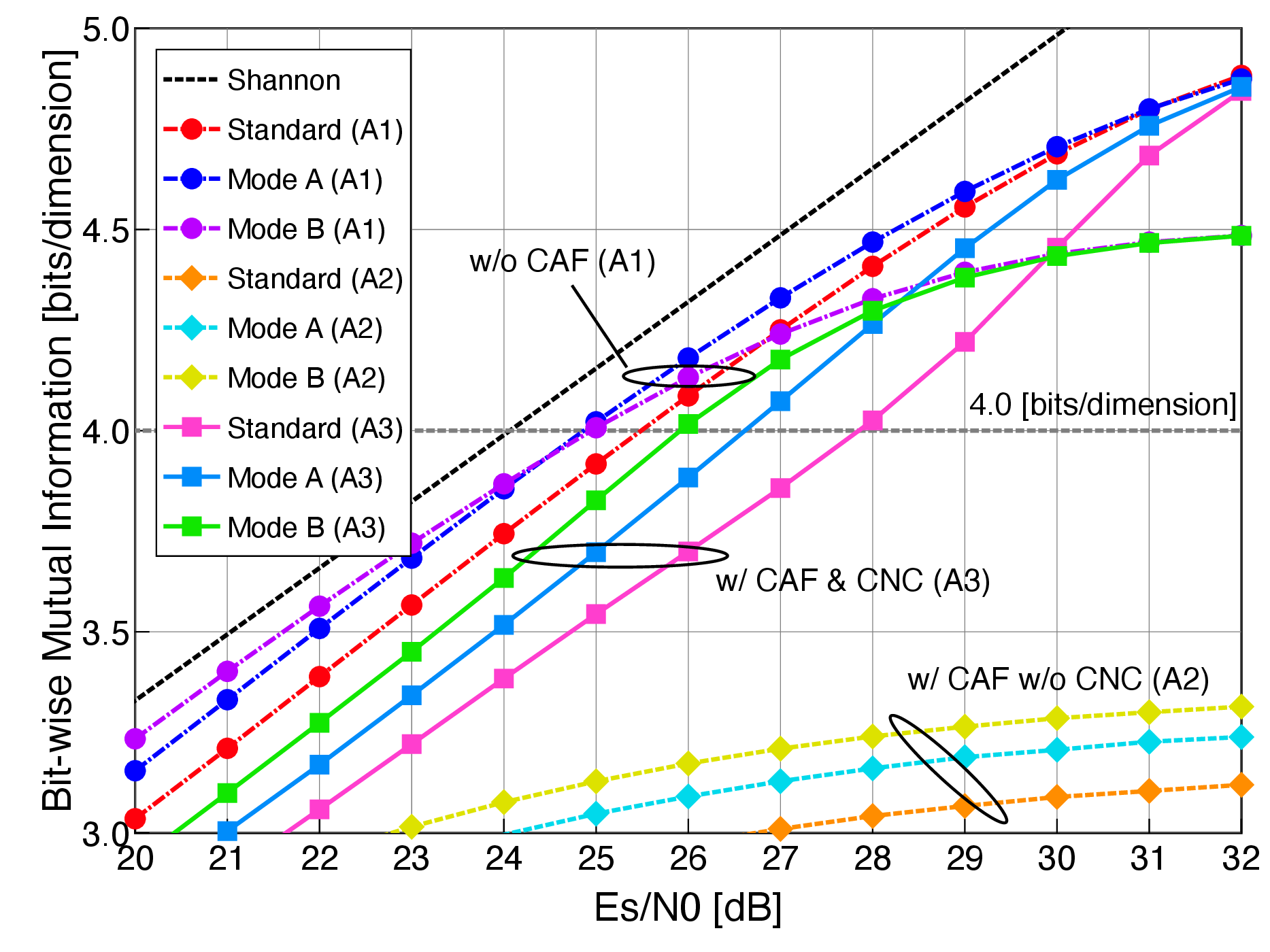}
	\caption{Comparison of BMI values for 32-PAM constellations over an AWGN channel. 
The following three systems are evaluated: {\bf A1}) 
Without CAF, {\bf A2}) with CAF only, and {\bf A3}) with CAF/CNC.}
    \label{fig:BMI}
\end{center}   
\vspace{-0.5em}
\end{figure}

\begin{figure}[tbp]
\begin{center}
	\includegraphics[width=\hsize]{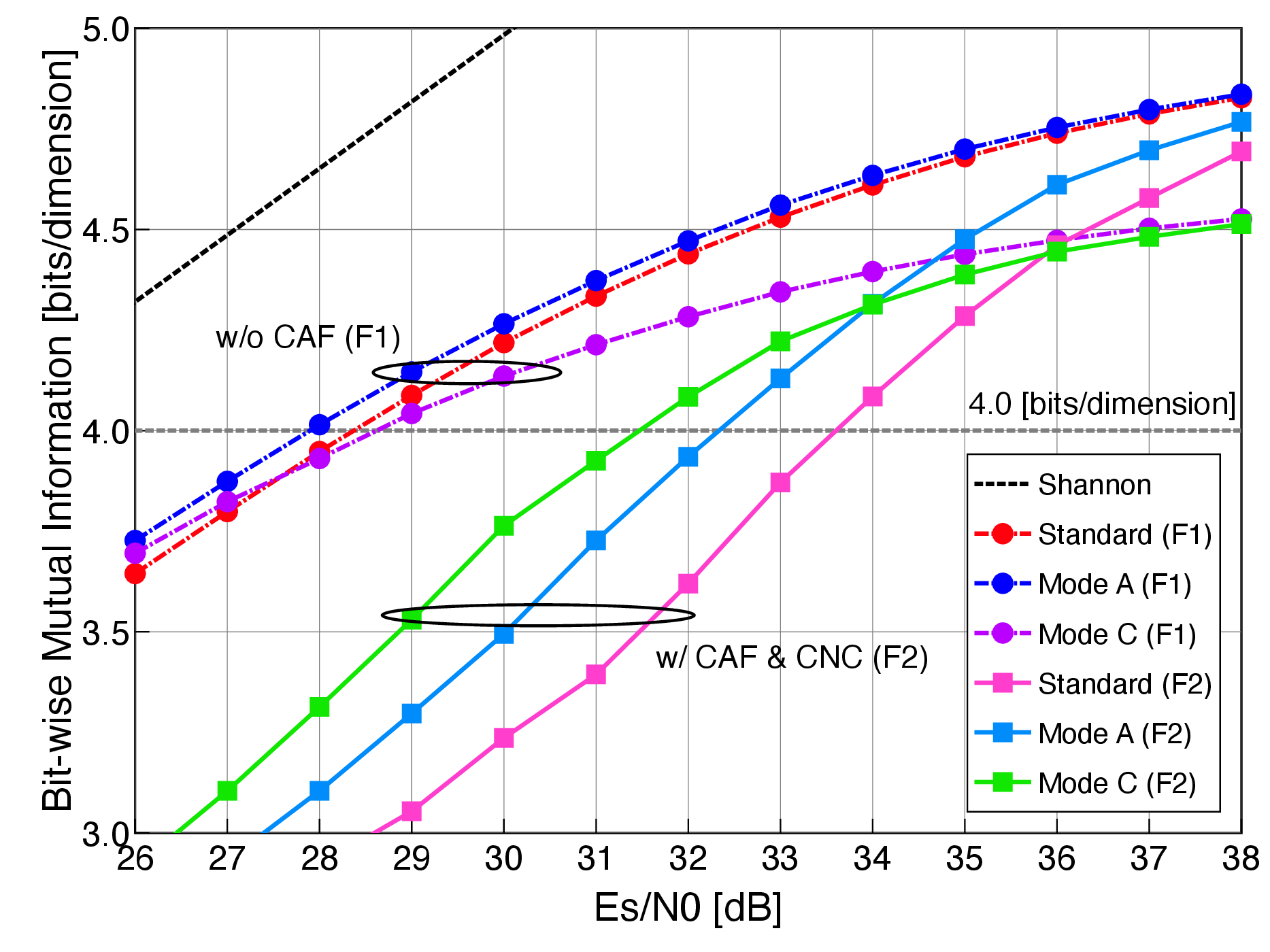}
	\caption{Comparison of BMI values for 32-PAM constellations over a
frequency-selective Rayleigh fading channel.
The following two systems are evaluated: {\bf F1}) Without CAF and {\bf F2}) 
with CAF and CNC. }
    \label{fig:BMIF}
\end{center}  
\vspace{-0.5em}
\end{figure}

For the purpose of investigating the effect of the parameter optimizations,
we evaluate the total BMI over an AWGN channel, focusing on 
the following three systems:
\begin{itemize}
\item[{\bf A1}:] BICM-OFDM without CAF (no PAPR reduction).
\item[{\bf A2}:] BICM-OFDM with CAF but without CNC.
\item[{\bf A3}:] BICM-OFDM with CAF and CNC.
\end{itemize}
The results are compared in Fig.~\ref{fig:BMI},
where 
the NENU constellations are fixed as either 
$\mathcal{X}_{\sf A}$ or $\mathcal{X}_{\sf B}$
over the entire channel SNR. The uniform PAM is also shown as a reference.
In {\bf A1}, 
the constellation $\mathcal{X}_{\sf A}$
demonstrates superior performance compared to the others
at the target information rate. 
On the other hand, 
in {\bf A2}, the achievable rates of all constellations 
significantly deviate from the target as a result of severe clipping noise.
Finally, in {\bf A3}, the information rate is successfully recovered by CNC, allowing the target information rate to be achieved with a lower PAPR compared to {\bf A1}.
The constellation $\mathcal{X}_{\sf B}$
exhibits superior performance, yielding a greater gain over 
the constellation $\mathcal{X}_{\sf A}$.
The results thus demonstrate the effectiveness of the proposed constellation design 
tailored to the BICM-OFDM system with CAF and CNC.

\subsubsection{Total BMI for Fading Channel}

We next consider 
the fading channel (5G NR TDL-B) for
the following two systems:
\begin{itemize}
\item[{\bf F1}:] BICM-OFDM without CAF (no PAPR reduction).
\item[{\bf F2}:] BICM-OFDM with CAF and CNC.
\end{itemize}
The BMI values are compared in Fig.~\ref{fig:BMIF},
where the parameters of the NENU constellations are fixed as either 
$\mathcal{X}_{\sf A}$ or $\mathcal{X}_{\sf C}$
over the entire channel SNR, along with the uniform PAM as a reference.
It is observed that for {\bf F1}, 
the gains achieved by the proposed constellations 
slightly
decrease 
compared to the results over the AWGN channel. 
This may be due to the fact that a single modulation scheme is applied across all subcarriers, while they experience different effective SNRs. 
Nevertheless, in {\bf F2}, significant gains by the proposed shaping 
are still observed when compared at the target information rate.
Additionally, the constellation $\mathcal{X}_{\sf C}$
(optimized for this system over a fading channel)
offers enhanced performance over $\mathcal{X}_{\sf A}$.

\section{Simulation Results}
\label{sec:Result}

In the remainder of this paper,
we present the system-level performance of
the BICM-OFDM employing the proposed NENU constellations in the presence of CAF and CNC
and make comparisons with other systems of similar transmitter complexity.
The system parameters are chosen identically to Table~\ref{tab:sim_setup_opt}.
As performance measures,
we evaluate the distribution of PAPR as well as BER.
We thus compare the following three systems: 
\begin{itemize}
\item[{\bf S1}:] Conventional BICM-OFDM without PAPR reduction.
\item[{\bf S2}:] BICM-OFDM employing CAF and CNC ($\gamma_\text{CR} = 1.5$).
\item[{\bf S3}:] BICM-OFDM employing DFT precoding.
\end{itemize}
Note that {\bf S3} is another low-complexity PAPR reduction scheme
comparable to CAF, and it is equivalent to a single-carrier system 
with a rectangular pulse shaping filter~\cite{myung2006single}. Therefore, its PAPR 
directly depends
on the underlying constellations~\cite{ochiai_wcl2012}.
For {\bf S3}, a conventional minimum mean square error~(MMSE) detector is employed at the receiver in the frequency-selective fading channel.
For each system, 
the following
modulation schemes are evaluated:
\begin{itemize}
\item The standard uniform $1024$-QAM.
\item The proposed 1024-ary NENU constellations
($\mathcal{X}_{\sf A}$, $\mathcal{X}_{\sf B}$, or $\mathcal{X}_{\sf C}$)
selected according to the system configuration.
\item The one-dimensional~(1D) $1024$-NUC of ATSC~3.0.
\item The two-dimensional~(2D) $1024$-NUC from~\cite{sillekens2022high}.
\end{itemize}
The 1D-NUC and 2D-NUC are optimized according to their respective design algorithms for the same target spectral efficiency of $8.0$ bits per complex channel use
and thus are suitable for the conventional OFDM system without CAF and CNC (i.e., {\bf S1} and {\bf S3}).

\subsection{CCDF of PAPR}

\begin{figure}[tbp]
\begin{center}
    \includegraphics[width=\hsize]{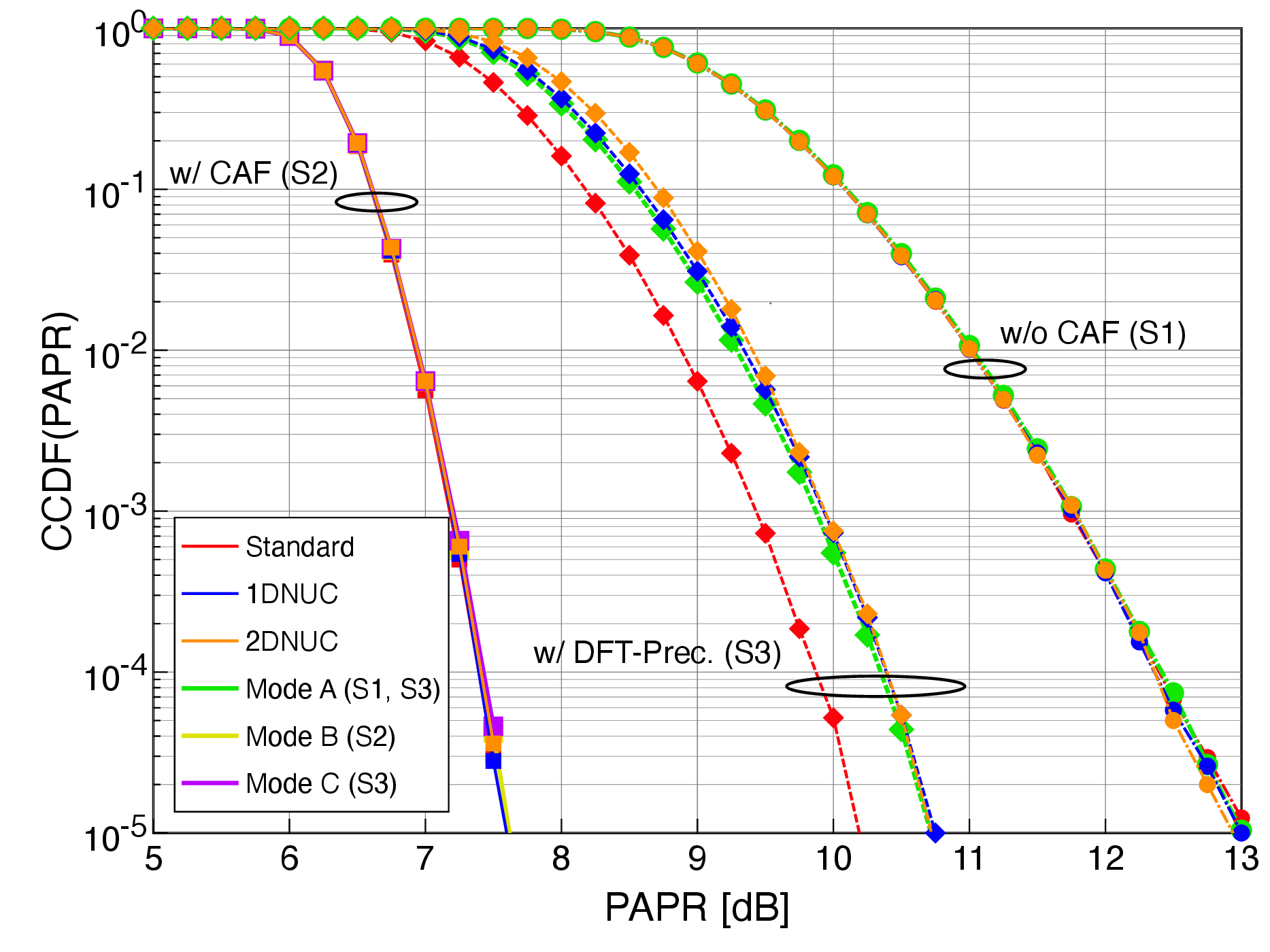}
    \caption{
Comparison of the CCDF of PAPR among the three systems: 
conventional OFDM without CAF, OFDM with CAF, and OFDM with DFT precoding (single-carrier), each employing the four different modulation schemes.}
    \label{fig:CCDF}
\end{center}
\vspace{-0.5em}
\end{figure}

The CCDF curves of the PAPR
among the three systems are shown in Fig.~\ref{fig:CCDF}, where all the aforementioned $1024$-ary constellations are compared. 
We observe no difference among the four curves plotted for {\bf S1} and {\bf S2}, suggesting that 
each subcarrier constellation does not affect PAPR distribution for OFDM. 
On the other hand, in {\bf S3}, the selection of constellations significantly affects the dynamic range of the transmit signals,
indicating that the use of constellation shaping increases the PAPR as expected~\cite{8227810},
even though it is still less than the conventional OFDM ({\bf S1}).
In contrast, the results for {\bf S2} indicate that 
CAF effectively reduces the peak power of OFDM signals
even in the presence of constellation shaping,
achieving a PAPR reduction of $5.0$ dB compared to {\bf S1} when measured at a CCDF of~$10^{-4}$. 

\subsection{BER Comparison Over AWGN Channel}

\begin{figure}[tbp]
\begin{center}
    \includegraphics[width=\hsize]{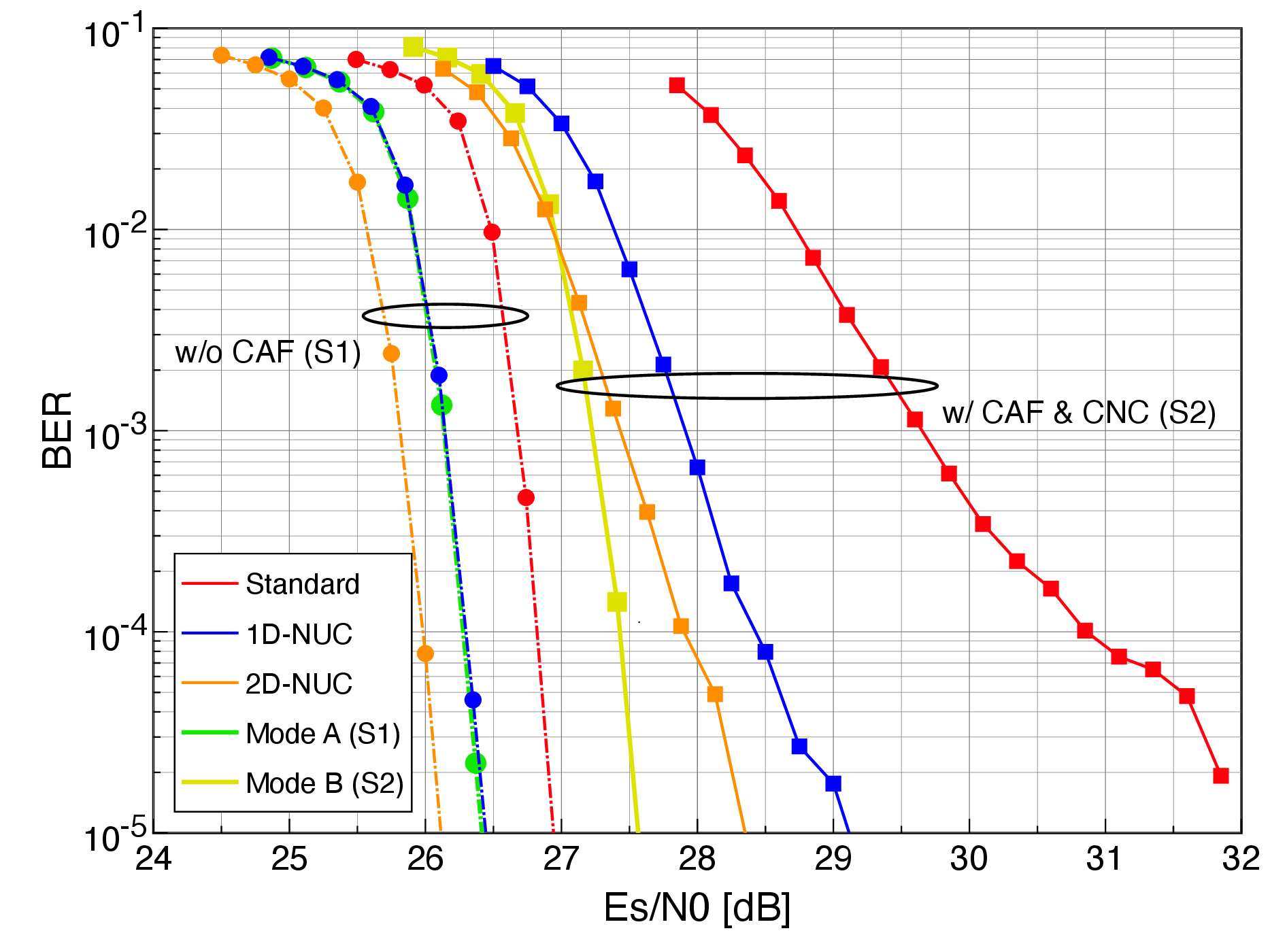}
    \caption{
BER comparison over an AWGN channel for the BICM-OFDM systems 
without CAF (no PAPR reduction) and with CAF/CNC employing the four different modulation schemes.}
    \label{fig:BER}
\end{center}
\vspace{-0.5em}
\end{figure}

Fig.~\ref{fig:BER} compares the BER performance over an AWGN channel for the BICM-OFDM systems using different modulation schemes, 
considering both cases without CAF (i.e., no PAPR reduction) and with CAF and CNC.
As for the proposed 
shaping scheme, 
the constellations 
${\mathcal{X}_{\sf A}}$ and ${\mathcal{X}_{\sf B}}$ are applied to ${\bf S1}$ and ${\bf S2}$,
respectively.
Note that the results of {\bf S3} are omitted as they exhibit similar performance
to {\bf S1} over the AWGN channel.
In the case of {\bf S1},
it can be confirmed that the performance gain achieved by 
the constellation ${\mathcal{X}_{\sf A}}$ 
agrees with that estimated from the BMI values in Fig.~\ref{fig:BMI}. 
The resulting BER performance is also comparable to that obtained by the one-dimensional NUC adopted in  ATSC~3.0.
Although the two-dimensional NUC proposed in\cite{sillekens2022high} achieves the largest shaping gain,
it requires much more demapping complexity compared with other modulation schemes based on one-dimensional constellations, as discussed in Section~III.

On the other hand, in the case of {\bf S2},
the constellation ${\mathcal{X}_{\sf B}}$ 
achieves a significant gain
over the standard uniform PAM.
Since ${\mathcal{X}_{\sf B}}$ is optimized 
in the presence of CAF/CNC, the BER exhibits a steep decay with increasing SNR, indicating strong robustness to residual clipping noise.
As a result, ${\mathcal{X}_{\sf B}}$ 
also outperforms both the one-dimensional NUC and the two-dimensional NUC, which are optimized for conventional OFDM systems without PAPR reduction, particularly at low BER levels.
These results confirm the effectiveness of optimizing the constellation for the specific OFDM system incorporating nonlinear CAF and CNC operations.
The shaping gain achieved by ${\mathcal{X}_{\sf B}}$,
assisted by the performance recovery through CNC, 
significantly reduces the gap relative to {\bf S1} (without PAPR reduction).
It is noteworthy that the performance gap between {\bf S1} with the uniform PAM
and {\bf S2} with ${\mathcal{X}_{\sf B}}$ is only about $0.50$ dB, while they have a 
notable
difference in terms 
of PAPR. This suggests that the proposed NENU constellation together with CAF and CNC
can significantly reduce PAPR without a noticeable SNR penalty compared to the conventional BICM-OFDM 
system.

\subsection{BER Comparison Over Fading Channel}

\begin{figure}[tbp]
\begin{center}
    \includegraphics[width=\hsize]{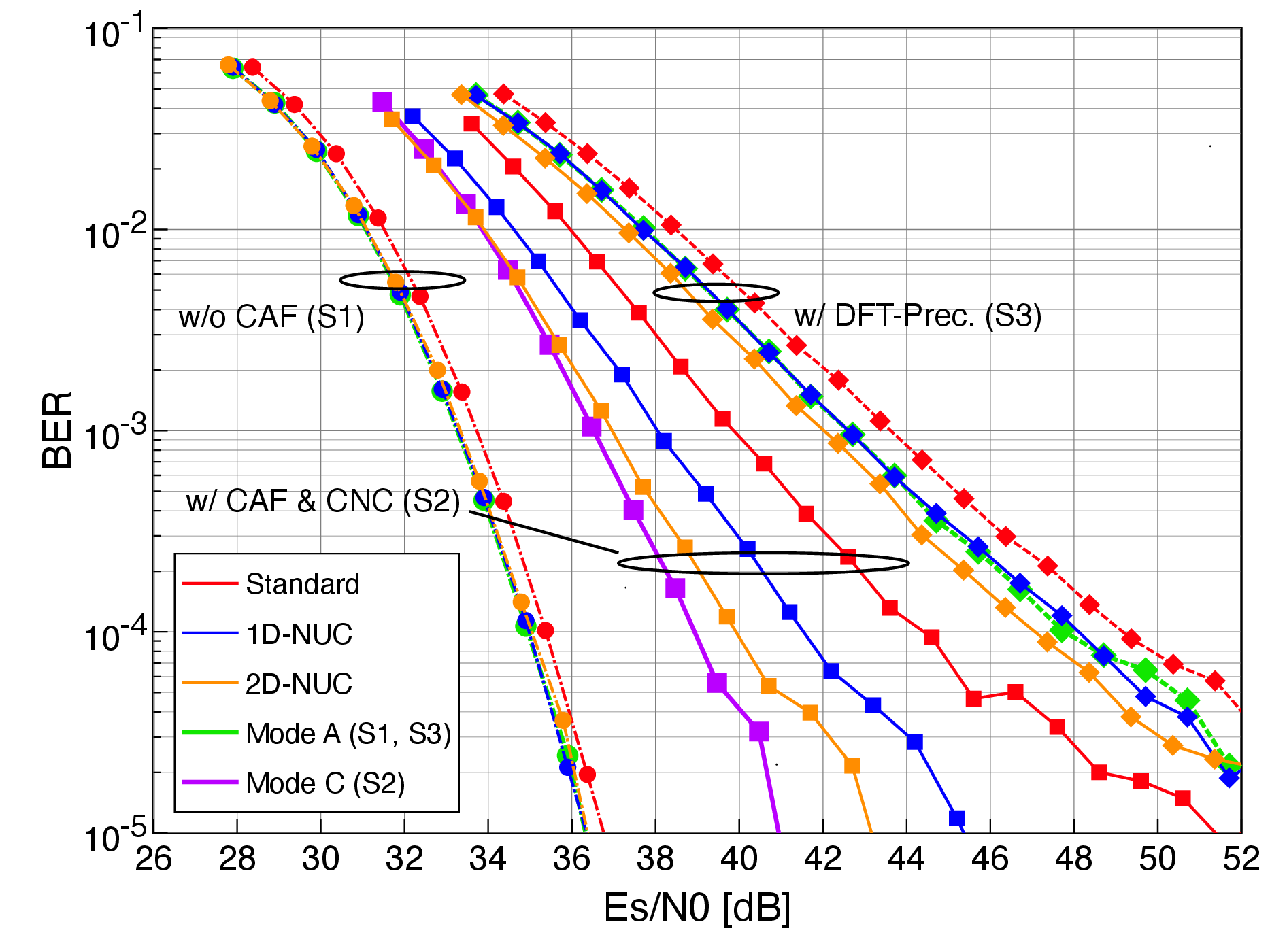}
    \caption{
BER comparison over  
the fading channel for the BICM-OFDM systems without CAF (no PAPR reduction), with CAF/CNC, and with DFT precoding, each employing the four different modulation schemes.}
    \label{fig:BERF}
\end{center}
\vspace{-0.5em}
\end{figure}

Finally, we compare BER performance among the three systems {\bf S1},
{\bf S2}, and {\bf S3}
over the 
fading channel (5G NR TDL-B) in Fig.~\ref{fig:BERF}. 
As for the proposed shaping scheme, 
the constellations
${\mathcal{X}_{\sf A}}$ is applied to both ${\bf S1}$ and ${\bf S3}$,
whereas ${\mathcal{X}_{\sf C}}$ is applied to ${\bf S2}$.
It is confirmed that the proposed NENU constellation provides a shaping gain for all systems over the practical fading channel as well.
Moreover, 
the performance of {\bf S3} (employing an MMSE detector)
is considerably inferior to that of the other OFDM-based systems. 
It is known that
single-carrier signals experience BER degradation 
when suboptimal detectors, such as MMSE, are employed
at the receiver~\cite{nasri2009performance,tajer2010diversity}. 
The results thus indicate that the performance loss associated with DFT precoding is 
more pronounced than that caused by CAF with CNC. 
Consequently, for PAPR reduction in OFDM systems utilizing
constellation shaping,  CAF combined with CNC is more advantageous than
DFT precoding 
from the viewpoint of PAPR as well as BER performance over fading channels.

\section{Conclusion}
\label{sec:Conclusion}

In this work, we developed high-capacity and low-PAPR
OFDM-BICM systems based on one-dimensional NENU constellations 
for shaping gain along with CAF and CNC for efficient signal peak power reduction
without noticeable performance degradation. As a result, the developed systems
enjoy significantly lower PAPR while achieving comparable performance with the standard
PAM-based BICM-OFDM system. 
The simulation results have demonstrated that 
significant performance gain can be achieved over the conventional BICM-OFDM with
DFT precoding in terms of both PAPR distribution and BER.
In our simulations, we have restricted our attention to the design of $32$-PAM (i.e., $1024$-QAM),
but its extension to even higher-order constellations targeting even higher information rates is
straightforward, considering the simplicity of the optimization process based on only the two shaping parameters.

As final remarks, our constellation design has been optimized for the simulated systems, and its robustness against other channels or system setups should be worth investigating. Furthermore, it is practically important to optimize the overall system by considering the effect of power amplifier nonlinearity in the presence of CAF/CNC, which we leave as future work.

\bibliographystyle{IEEEtran}
\bibliography{main}

\end{document}